\documentclass[12pt,tightenlines,eqsecnum,floats,showpacs,nofootinbib,amsmath,amssymb,aps,prd]{revtex4}
\usepackage{amsmath,amssymb}
\usepackage{amsfonts,amssymb,mathrsfs}
\usepackage[bookmarks=false,pdfstartview=FitH]{hyperref}

\def\be{\begin{equation}}
\def\ee{\end{equation}}
\def\nn{\nonumber}
\def\f{\frac}
\def\tf{\tfrac}
\def\sgn{{\rm sgn}}
\def\pl{{\rm Pl}}
\def\lp{\ell_\pl}
\def\b{\bar}
\def\d{\dot}
\def\h{\hat}
\def\t{\tilde}

\def\wh{\widehat}

\def\bra{\langle}
\def\ket{\rangle}
\def\dd{{\rm d}}

\def\de{\delta}

\def\ga{\gamma}
\def\la{\lambda}
\def\om{\omega}

\def\De{\Delta}

\def\oe{\mathring{e}}
\def\ow{\mathring{\omega}}
\def\oq{\mathring{q}}
\def\oep{\mathring{\epsilon}}
\def\oV{\mathring{V}}
\def\lo{\ell_o}
\def\mH{\mathcal{H}}
\def\mC{\mathcal{C}}
\def\mO{\mathcal{O}}
\def\mV{\mathcal{V}}
\def\ba{\begin{eqnarray}}
\def\ea{\end{eqnarray}}

\begin{document}

\pagestyle{plain}

\title{Quantization Ambiguities and Bounds on Geometric Scalars\\ in Anisotropic Loop Quantum Cosmology}

\author{Parampreet Singh} \email{psingh@phys.lsu.edu}
\affiliation{Department of Physics and Astronomy,
Louisiana State University, Baton Rouge, 70803}

\author{Edward Wilson-Ewing} \email{wilson-ewing@phys.lsu.edu}
\affiliation{Department of Physics and Astronomy,
Louisiana State University, Baton Rouge, 70803}

\begin{abstract}

We study quantization ambiguities in loop quantum cosmology
that arise for space-times with non-zero spatial curvature and
anisotropies.  Motivated by lessons from different possible loop
quantizations of the closed Friedmann-Lema\^itre-Robertson-Walker
cosmology, we find that using open holonomies of the extrinsic curvature,
which due to gauge-fixing can be treated as a connection,
leads to the same quantum geometry effects that are found in
spatially flat cosmologies.  More specifically, in contrast to
the quantization based on open holonomies of the Ashtekar-Barbero
connection, the expansion and shear scalars in the effective theories of
the Bianchi type II and Bianchi type IX models have upper bounds, and these
are in exact agreement with the bounds found in the effective theories of the
Friedmann-Lema\^itre-Robertson-Walker and Bianchi type I models in loop
quantum cosmology.   We also comment on some ambiguities present in
the definition of inverse triad operators and their role.

\end{abstract}


\maketitle

\section{Introduction}
\label{s.intro}

Loop quantum cosmology (LQC) is a quantization of cosmological space-times using
the techniques of loop quantum gravity (LQG) \cite{Ashtekar:2011ni,Bojowald:2008zzb,
Banerjee:2011qu,Agullo:2013dla}. One starts by exploiting the underlying symmetries
of the spatial manifold in order to symmetry reduce the Ashtekar-Barbero connection
$A^i_a$ and the densitized triad $E^a_i$, and then takes the holonomies of the
connection and fluxes of the densitized triads as the elementary variables.  The
classical Hamiltonian constraint is expressed in terms of these elementary variables
and quantized.  The resulting physical evolution turns out to be strikingly different
from the Wheeler-DeWitt theory: it has been demonstrated in various models that the
cosmological singularity is avoided, and replaced by a quantum bounce. The existence
of a quantum bounce is a direct consequence of the underlying quantum geometry in LQC
which manifests itself when the space-time curvature approaches the Planck scale.  On
the other hand, in the regime where the space-time curvature is weak, the dynamics of
LQC are well approximated by the classical theory.

The inputs of quantum geometry occur in two types of terms that appear in the Hamiltonian
constraint.  These are the field strength of the connection, and terms involving inverse
powers of triads.  These two types of terms are handled in different ways and there are
subtleties in each of their implementations, so it is useful to review how the relevant
operators are defined in LQC.

Let us begin by considering inverse triad operators.  There is not
only a considerable ambiguity in the definition of the inverse volume operator
\cite{Bojowald:2004xq}, but it is also unclear whether inverse volume corrections
should be implemented in the same manner for both the gravitational and matter
contributions to the scalar constraint, or whether the nature of the
inverse volume operators depends on the setting they appear in. Furthermore, even in
the manner of their implementation in the matter part there is an additional ambiguity
regarding whether one should combine the energy density and the square root of the
spatial metric before defining the inverse triad operators or not \cite{Singh:2005km}.
While inverse triad effects turn out to be insignificant for the dynamics of isotropic
models (in fact they vanish for spatially non-compact models \cite{Ashtekar:2006wn})
when compared to the effects coming from the field strength operator, they do lead
to a rich phenomenology, see e.g.\ \cite{Singh:2003au}, and have been shown to be
important in the spatially curved anisotropic models \cite{Gupt:2011jh, Corichi:2009pp}.
There has also been some effort in trying to obtain inverse triad operators that do
not vanish in non-compact spaces \cite{Bojowald:2011iq}; while no such satisfactory
operator has been found so far, there do exist some semi-classical expressions which
may capture some of the correct physics.  One of the goals of this paper is to
parametrize some of the ambiguities present in the definition of inverse triad
operators in LQC and see whether some of them can be resolved.

The operator corresponding to the field strength of the Ashtekar-Barbero connection
is obtained by following the strategy that is used in the gauge theories, which we
will call the `F' loop quantization procedure. One writes the field strength in
terms of holonomies of the connection over a closed loop, generally taken to be
a square whose edges are generated by the fiducial triads on the spatial manifold,
and the loop is shrunk to the area equal to the minimum eigenvalue of the area
operator in LQG.  While this procedure is straightforward for the spatially flat
isotropic Friedmann-Lema\^itre-Robertson-Walker (FLRW) cosmology \cite{Ashtekar:2003hd,
Ashtekar:2006wn} as well as the spatially flat anisotropic Bianchi I model
\cite{Ashtekar:2009vc}, it requires more care when applied to the spatially
closed FLRW model \cite{Ashtekar:2006es,Szulc:2006ep}.  Nonetheless, in these cases
the resulting expression of the field strength is expressed in terms of almost
periodic functions of connection, which are easy to quantize.  However, this
procedure does not yield almost periodic functions of the connection for
the spatially open FLRW model or for spatially curved anisotropic models;
since the resulting functions are not almost periodic, it is not known
how to quantize them.  To overcome this problem, two alternatives have been
suggested in the literature, both of which propose taking parallel transport
along open edges, rather than around a closed loop.  The first suggestion is
called the `K' quantization, where the extrinsic curvature $K_a^i$ is expressed
in terms of its parallel transport along open edges that do not form a closed
loop \cite{Bojowald:2003mc,Bojowald:2003xf,Vandersloot,Vandersloot:2006ws};
this allows a treatment of the spatially curved FLRW and Bianchi models.
(Note that in this setting, due to gauge-fixing, the extrinsic curvature
can be treated as a connection \cite{Vandersloot:2006ws}.)
The second possibility, called the `A' quantization, is to do the same but with the
Ashtekar-Barbero connection rather than the extrinsic curvature, this leads to a
successful quantization of the Bianchi II and Bianchi IX models \cite{Ashtekar:2009um,
WilsonEwing:2010rh} and was also used in order to obtain an alternative quantization
of the spatially closed FLRW model \cite{Corichi:2011pg,Corichi:2013usa}.  As the
`A' procedure is based on holonomies of the Ashtekar-Barbero connection rather than
the parallel transport of the extrinsic curvature, it is often believed to be a better
alternative than the `K' loop quantization.

It is important to point out that in the spatially flat cosmologies, namely the flat
FLRW and Bianchi I models, the `F', `A' and `K' loop quantizations can all be
performed and are in fact equivalent.  For the closed FLRW cosmology, once again
all three of the `F', `A' and `K' loop quantizations can be performed, but the
resulting quantum theories in this case are in fact inequivalent.  In other
cosmologies with spatial curvature, only the `A' and `K' loop quantizations are
viable, and the resulting quantum theories are inequivalent.

A particularly nice property of LQC models is that states that are initially sharply
peaked remain sharply peaked at all times, even during and beyond the bounce
\cite{Corichi:2007am, Kaminski:2010yz, Corichi:2011rt}.  Thus, for sharply peaked
states it is possible to speak of an effective geometry, whose dynamics are given by
an effective Hamiltonian constraint \cite{Willis,Taveras:2008ke}%
\footnote{For an alternate approach to the effective theory, see
Ref.\ \cite{Bojowald:2012xy}. For a discussion of the differences
in the underlying assumptions of the different ways to derive an
effective Hamiltonian in LQC, see Ref.\ \cite{Ashtekar:2011ni}.}.
Using numerical simulations, a comparison of the full quantum dynamics and the
effective dynamics generated by the effective Hamiltonian constraint show that the
two are in good agreement, including at the bounce point \cite{Ashtekar:2006wn}.
It has been argued that the modifications to the Friedmann equation which do not
depend on the properties of the state capture the underlying quantum evolution
to a good approximation so long as the volume of the universe remains large
compared to the Planck volume \cite{psvt}.  It has further been argued that the
state-dependent terms due to quantum fluctuations have a negligible effect on
the effective trajectory derived from the modified Friedmann equation also so
long as the volume of the universe remains large compared to the Planck volume
\cite{Rovelli:2013zaa}.  Since the effects of the field strength and inverse
triad operators are included in the effective Hamiltonian constraint in the form
of correction functions, it is possible to study their ambiguities already at the
effective level.  As it is particularly easy to study quantum gravity effects in
this manner, we will mostly work in the effective theory in this paper.

Amongst the three space-times for which the `F' quantization is possible ---the flat
and closed FLRW and the Bianchi I models--- the resulting effective theories share
several properties in the Planck regime.  Perhaps the most important of these
is the fact that the expansion scalar and the energy density of
the matter field are both bounded above.  In fact, for universes that remain much
larger than the Planck volume at all times, the upper bound on the expansion scalar
is the same for all three cosmologies as is the bound on the energy density
\cite{Ashtekar:2009vc,Corichi:2009pp,Gupt:2011jh}.

This harmony is broken for the `A' quantization of the Bianchi II and Bianchi IX
models.  In the effective space-time description, the expansion and shear scalars
as well as the energy density of the Bianchi II cosmology turn out to be unbounded
unless the matter satisfies the weak energy condition \cite{Gupt:2011jh}.
However, the bounds on these quantities turn out to be significantly different than
the bounds obtained for the FLRW and Bianchi I models in LQC.  The situation for
the Bianchi IX model is even more problematic.  For an isotropic approach to
the singularity, the expansion and shear scalars are bounded only when inverse
triad effects are included.  If the approach to the singularities is anisotropic,
then these scalars are unbounded even when inverse triad corrections are included
\cite{Gupt:2011jh}.  Finally, unlike in the Bianchi II model, the energy density
in Bianchi IX is unbounded, unless one defines the energy density by including inverse
triad modifications in the matter Hamiltonian and multiplying it with an inverse
volume operator as suggested in \cite{Singh:2005km}.  Therefore, we see that the
`A' loop quantization of the Bianchi models does not give the same qualitative
predictions as the standard `F' loop quantization of simpler cosmologies.

As mentioned earlier, it is possible to study the effective equations for the closed
FLRW space-time, where the `F', `A' and `K' loop quantizations are all possible.
As we shall show in Sec.\ \ref{ss.comp}, the qualitative predictions of the
`F' and `A' effective theories are significantly different, and this raises
important objections to the validity of the `A' loop quantization as an
approximation to the standard `F' loop quantization.  On the other hand, we
find that the effective theory of the `K' loop quantization in fact {\it does
capture} the same qualitative predictions of the `F' quantization, and so
the `K' prescription appears to be a good approximation to the usual loop
quantization.  Another of the main goals of this paper is to compare the
effective theories of the `A' and `K' loop quantizations in richer models
in order to determine which quantization better captures the physics
observed in other LQC models.

Concerning the effective theory, our main focus in this paper will be on the
geometric quantities of the expansion and shear scalars and therefore it is
important to recall their geometric interpretations.  These scalars are
particularly important as they play a crucial role in understanding the
resolution of singularities.  The expansion scalar is defined via the
trace of $\nabla_b \xi_a$ where $\xi_a$ corresponds to a congruence of
cosmological observers defined with respect to proper time (i.e., lapse
$N=1$).  It is related to the trace of the extrinsic curvature $K_{ab}$
as $\theta = K$, and in the cosmological setting to the (mean) Hubble
rate as $\theta = 3 H$. In the classical theory, through the Raychaudhuri
equation, the evolution of the expansion scalar is related to the
anisotropic shear (which vanishes in FLRW space-times) and components
of the Ricci tensor (assuming no vorticity in the matter fields).  As a
curvature singularity is approached, the expansion and shear scalars diverge
and geodesic evolution breaks down.  On the other hand, if the expansion
and shear scalars can be shown to remain bounded at all times, then this is
a strong indication that there are no geodesic singularities.
Therefore, it is of interest to study the expansion and shear scalars in
LQC in the high curvature regime where they diverge classically in order to
see whether they remain bounded once quantum geometry effects have been
included.  As mentioned above, this analysis has been performed for the
flat FLRW and Bianchi I models in LQC, with the result that both the
expansion and shear are bounded above \cite{Corichi:2009pp}, and it can
be shown that the effective space-time is free of geodesic singularities
\cite{Singh:2009mz, Singh:2011gp}.  Finally note that it is
not necessary to bound the matter energy density in order to show the
absence of geodesic singularities; for this reason a bound on
the expansion and shear is more important than a bound on the energy
density.

In this paper we study the quantization ambiguities of the field strength
and inverse triad operators in LQC, working in what is arguably the richest
homogeneous space-time, the Bianchi IX universe.  We start by comparing the
`A' and `K' loop quantization procedures, first in the closed FLRW space-time
---the isotropic limit of Bianchi IX--- in Sec.\ \ref{s.closed}, and then
in the Bianchi IX model itself in Sec.\ \ref{s.b-ix}; we compare the two
quantization prescriptions via the predictions of their respective effective
theories concerning the geometric quantities of the expansion and shear scalars.
Then in Sec.\ \ref{s.inv}, we parametrize the ambiguities that appear in the
possible definitions of the inverse volume operators and we also propose an
alternative inverse volume operator which is free of any ambiguities.
Finally, Appendix \ref{s.b-ii} is devoted to the Bianchi II model, where
we derive its `K' loop quantization and also determine the expansion
and shear scalars in the effective theory.

\section{The Closed FLRW Space-time}
\label{s.closed}

In this section, we briefly review the loop quantization of the closed
FLRW model with a massless scalar field performed in \cite{Ashtekar:2006es}.
We begin with an outline of the classical framework in Sec.\ \ref{ss.class-closed}
and then briefly discuss the quantization of this model using holonomies of
the $SU(2)$ connection $A^i_a$ in Sec.\ \ref{ss.closed-standard}.

The quantization gives a quantum difference equation with a uniform
discretization in volume which results in a quantum bounce when the energy
density of the scalar field becomes approximately $\rho \approx 0.41
\rho_{\mathrm{Pl}}$.  The bounce is nicely captured by the dynamics
generated by an effective Hamiltonian description which reveals an
upper bound on the value of the expansion scalar in this model.  In
Sec.\ \ref{ss.comp}, we contrast three different effective Hamiltonian
descriptions for the three different ways to loop quantize this model: (i) the
`F' loop quantization given in \cite{Ashtekar:2006es} and reviewed in
Sec.\ \ref{ss.closed-standard}, (ii) the `A' loop quantization obtained
by polymerizing the connection $A^i_a$ given in \cite{Corichi:2011pg},
and, (iii) the `K' loop quantization where the extrinsic curvature $K^i_a$
is polymerized.  We show that the expansion is bounded in the effective
theories for the `F' and `K' ---but not for the `A'--- loop quantizations
of the closed FLRW model.

\subsection{The Classical Space-time}
\label{ss.class-closed}

As the topology of the closed FLRW universe is $\mathbb{S}^3$, it is useful
to introduce a basis, called fiducial triads, for the 3-sphere.
Setting the radius of the 3-sphere to be 2, a possible choice is
\begin{align}
\label{tri-1}
\oe^a_1 &= \f{\sin \ga}{\sin \beta} \left( \f{\partial}{\partial \alpha} \right)^a
- \cos \ga \left( \f{\partial}{\partial \beta} \right)^a
+ \f{\cos \beta \sin \ga}{\sin\beta} \left( \f{\partial}{\partial \ga} \right)^a, \\
\oe^a_2 &= -\f{\cos \ga}{\cos \beta} \left( \f{\partial}{\partial \alpha} \right)^a
+ \sin \ga \left( \f{\partial}{\partial \beta} \right)^a
+ \f{\cos \beta \cos \ga}{\sin\beta} \left( \f{\partial}{\partial \ga} \right)^a, \\
\label{tri-3}
\oe^a_3 &= \left( \f{\partial}{\partial \ga} \right)^a,
\end{align}
where $\alpha \in [0, 2\pi), \beta \in [0, \pi]$ and $\ga \in [0, 4\pi)$.
Although these coordinates become singular at the poles $\beta = 0$ and
$\beta = \pi$, they hold elsewhere on the 3-sphere.

The fiducial triads for the 3-sphere satisfy the usual $SO(3)$ commutation relations,
\be
[\oe_i, \oe_j] = -\oep_{ij}{}^k \oe_k,
\ee
where $\oep_{ijk}$ is totally antisymmetric and $\oep_{123} = 1$.  Note
that there is a choice to be made regarding the overall sign appearing
in this commutation relation, here we follow the conventions used in
\cite{WilsonEwing:2010rh} (the opposite convention is used
in \cite{Ashtekar:2006es, Bojowald:2003xf}).
While this choice affects the sign of the spin-connection, it does not change
the field strength of the spin-connection (which is what appears in the
scalar constraint).

The fiducial co-triads dual to the fiducial triads are
\begin{align}
\label{co-1}
\ow_a^1 &= \sin\beta \sin\ga \, (\dd\alpha)_a + \cos\ga \, (\dd \beta)_a, \\
\ow_a^2 &= -\sin\beta \cos\ga \, (\dd\alpha)_a + \sin\ga \, (\dd \beta)_a, \\
\label{co-3}
\ow_a^3 &= \cos\beta \, (\dd\alpha)_a + (\dd\ga)_a,
\end{align}
and the fiducial spatial metric is given by
$\oq_{ab} = \ow_a^i \ow_{bi}$.
From this, it follows that the determinant of the fiducial metric is
$\oq = \sin^2 \beta$, and it is easy to check that the volume of a 3-sphere
with a radius of $r_o = 2$ is given by $\oV = 2 \pi^2 r_o^3 = 16 \pi^2$.
It is also useful at this point to introduce $\lo = \oV^{1/3}$.

The only degree of freedom in an FLRW universe is the scale factor
$a(t)$, which is the factor of proportionality between the fiducial
co-triad and the physical co-triad,
\be
\om_a^i = a(t) \, \ow_a^i,
\ee
and then the metric for the closed FLRW universe is given by
\be
q_{ab} = \om_a^i \om_{bi} = a(t)^2 \, \oq_{ab}.
\ee

In LQG, the fundamental variables are fluxes of the densitized triads
$E^a_i = \sqrt{q} e^a_i$ and holonomies of the Ashtekar-Barbero connection
$A_a^i$.  The densitized triads and Ashtekar-Barbero connection can be
parametrized by
\be
E^a_i = \f{p}{\lo^2} \, \sqrt{\oq} \, \oe^a_i, \qquad
A_a^i = \f{c}{\lo} \, \ow_a^i,
\ee
where the variable $p$ is related to the scale factor by
$|p| = a^2 \lo^2$ (the sign of $p$ determines whether the
triads $e^a_i$ are right- or left-handed).  The Poisson bracket
of the basic variables is simply
\be
\{c, p\} = \f{8 \pi \ga G}{3}.
\ee

As the diffeomorphism and Gauss constraints are automatically satisfied
by this parametrization of $E^a_i$ and $A_a^i$, only the scalar constraint
is left,
\be \label{k1_cons}
\mC_H = \int \left[ \f{-N E^a_i E^b_j}{16 \pi G \ga^2 \sqrt{q}}
\epsilon^{ij}{}_k \Big( F_{ab}{}^k - (1 + \ga^2) \Omega_{ab}{}^k
\Big) + N \mH_{\rm matter} \right] \approx 0,
\ee
where $N$ is the lapse, $\ga$ is the Immirzi parameter (not to be confused
with the coordinate on the 3-sphere), while $F_{ab}{}^k$ and $\Omega_{ab}{}^k$
are the field strengths of $A_a^i$ and the spin-connection respectively.
Since we are mostly interested in the gravitational sector in this paper,
for the sake of simplicity we will work with a massless scalar field, in
which case $\mH_{\rm matter} = p_\phi^2 / 2 p^{3/2}$, with $p_\phi$ being
the momentum conjugate to the scalar field $\phi$.

The field strength $F_{ab}{}^k$ is given by
\be
F_{ab}{}^k = 2 \partial_{[a}^{} A_{b]}^k + \epsilon_{ij}{}^k A_a^i A_b^j
= \left( \f{c}{\lo} - \f{c^2}{\lo^2} \right) \epsilon_{ij}{}^k \ow_{[a}^i \ow_{b]}^j,
\ee
while the spin-connection $\Gamma_a^i$ is
\be
\Gamma_a^i = -\epsilon^{ijk} e^b_j \left( \partial_{[a} \om_{b]k} + \f{1}{2}
e^c_k \om_a^l \partial_{[c} \om_{b]l} \right) = -\f{1}{2} \ow_a^i,
\ee
which determines the spatial curvature,
\be
\Omega_{ab}{}^k = -\f{1}{4} \epsilon_{ij}{}^k \ow_{[a}^i \ow_{b]}^j.
\ee
From these results, it is easy to see that the scalar constraint for $N = 1$
is
\be\label{k1_class}
\mC_H = -\f{3 \sqrt{|p|}}{8 \pi G \ga^2} \left( c^2 - \lo c + \f{\lo^2 (1 + \ga^2)}{4}
\right) + \f{p_\phi^2}{2 \, p^{3/2}} \approx 0.
\ee
Then, by $\dot{\mO} = \{ \mO, \mC_H \}$,
\be
\dot{p} = \f{\sqrt{|p|}}{\ga} \left( 2 c - \lo \right),
\ee
and by using $\mC_H = 0$, the Friedmann equation is recovered,
\be
H^2 = \f{8 \pi G}{3} \rho - \f{1}{4 a^2}.
\ee
Here $H = \dot{p}/2p$ is the Hubble rate, and $\rho = p_\phi^2 / 2 |p|^3$
is the energy density due to the scalar field.  Note that the factor
of 4 in the denominator in the last term appears as the radius of the
3-sphere with respect to the fiducial triads is 2.

\subsection{Review of the Standard LQC Quantization}
\label{ss.closed-standard}

As in LQG, the elementary variables for quantization of the gravitational sector are the fluxes of the triad and the holonomies of the connection $A^i_a$. Due to homogeneity, the fluxes turn out to be proportional to the triad $p$, whose eigenvalues are given by
\be
\wh{\, p \,} |p\rangle = 
p |p\ket,
\ee
where the eigenkets satisfy $\langle p_1| p_2 \rangle = \delta_{p_1 , \, p_2}$, note that this is the Kronecker delta function, not the Dirac delta distribution.

Working in the $j=1/2$ representation of $SU(2)$, the operator corresponding to holonomies of the connection $A^i_a$ computed along straight edges of length $\mu \lo$ is
\be
h_k^{(\mu)} = \cos \f{\mu c}{2} \mathbb{I} + 2 \, \sin \f{\mu c}{2} \tau_k,
\ee
where $\mathbb{I}$ is the $2 \times 2$ identity matrix and the $\tau_i$ are a basis of the Lie algebra $\mathfrak{su}(2)$; this operator has the following action on the eigenkets $|p\rangle$:
\be
\hat h_k^{(\lambda)} |p\rangle = \f{1}{2} \Big(|p + \tf{8 \pi \ga \lp^2}{6} \mu\rangle + |p - \tf{8 \pi \ga \lp^2}{6} \mu \rangle\Big) \mathbb{I} - i \Big( |p + \tf{8 \pi \ga \lp^2}{6} \mu \rangle - |p - \tf{8 \pi \ga \lp^2}{6} \mu \rangle\Big) \tau_k ~.
\ee
The elements of the holonomies of the connection generate an algebra of almost periodic functions, and the kinematical Hilbert space in the loop quantization of the closed FLRW model is the space of
square integrable functions on the Bohr compactification of the real line: $L^2(\mathbb{R}_{\rm{Bohr}}, \dd \mu_{\rm{Bohr}})$.  On this space, a state $|\Psi\rangle$ can be expanded as a countable sum of the eigenstates of the triad operator.

To obtain the physical Hilbert space, as a first step, we express the classical Hamiltonian constraint in terms of the elementary variables -- the holonomies of the connection and the triad $p$ at the operator level. At the operator level, the term involving the product of triads in Eq.\ (\ref{k1_cons}) can be written as
\be
\epsilon_{ijk} \widehat{\f{E^{aj} E^{bk}}{{\rm det} \, e}} = \sum_k \f{\mathring{\epsilon}^{abc} \ow_c^k}{2 i \lo \gamma \pi G \hbar \lambda} \mathrm{Tr} \Big( \hat h_k^{(\lambda)} [h_k^{(\lambda) -1}, \hat V] \tau_i \Big),
\ee
where we have used $\sqrt{q} = {\rm det} \, e$. In order to express the field strength $F_{ab}{}^k$ in terms of holonomies we consider a square loop $\Box_{ij}$ constructed from left and right invariant vector fields on the manifold.  This gives \cite{Ashtekar:2006es}
\be
\widehat{F_{ab}{}^k} = \f{\epsilon_{ij}{}^k \ow^i_a \ow^j_b}{\bar \mu \lo^2} \left(\sin^2 \bar \mu\left(c - \f{\lo}{2} \right) - \sin^2\left(\f{\bar \mu \lo}{2}\right)\right),
\ee
where
\be \label{bar-mu}
\bar \mu^2 |p| = \Delta = 4 \sqrt{3} \pi \gamma \, \lp^2 ~,
\ee
the minimal non-zero eigenvalue of the area operator in LQG \cite{Ashtekar:2009vc}.
Since $\bar \mu$ is a function of the triad, the action of holonomies of the $\b\mu$ type
on the kets $|p\ket$ is no longer a simple translation. For this reason, a more convenient basis is provided by the eigenkets of the operator corresponding to the volume $V = |p|^{3/2}$,
\be
\wh V |v \rangle = 2 \pi \ga \lp^2 \sqrt\Delta \, |v| \, |v \rangle,
\ee
and then the action of elements of the holonomy algebra on the eigenkets $| v\rangle$ is simply given by a uniform translation in the volume operator,
\be
\widehat{e^{-i \bar \mu c/2}} \, |v \rangle  = |v + 1\rangle.
\ee

The matter part of the quantum Hamiltonian constraint results from the action of the
operator $\hat \mC_{\mathrm{matter}} = {\widehat{{p_\phi^2/2V^2}}}$ where the action of
the operator $\hat p_\phi$ is given by differentiation,
$\hat p_\phi \Psi(v,\phi) = - i \hbar \partial_\phi \Psi(v,\phi)$, while
the action of the inverse volume operator can be computed by adapting the Thiemann
inverse triad identities \cite{Thiemann:1996aw} to the reduced phase space of LQC.
We discuss inverse volume operators in considerable detail in  Sec.\ \ref{s.inv}
[this particular operator is derived in Eq.\ \eqref{bv2}], for now we shall simply quote
the result,
\be \label{bv1}
\wh{\, \f{1}{V} \,} |v\rangle = \f{|v|}{2 \pi \ga \lp^2 \sqrt\Delta}  \left(\f{3}{2}\right)^3
\Big| |v + 1|^{1/3} - |v - 1|^{1/3} \Big|^3 \, |v \rangle = B(v) |v\ket ~,
\ee
where we have introduced the function $B(v)$ as a shorthand for the eigenvalues
of the inverse volume operator.

Combining the terms of the gravitational and matter parts of the constraint,
the resulting quantum Hamiltonian constraint operator becomes
\ba \label{k1qc}
\partial_\phi^2 \Psi(v,\phi) &=& -\Theta \Psi(v,\phi) \nn \\
&=&  \pi G \sqrt\f{4 \pi \ga \lp^2}{3 \Delta} \cdot \f{1}{B(v)}
\Bigg[ \Big(C^+(v) \Psi(v + 4,\phi) + C^0(v) \Psi(v,\phi)
+ C^-(v) \Psi(v-4,\phi) \Big) \nn \\
&& + \Bigg( |v| \bigg(\sin^2 \f{\lo\sqrt\Delta}{2\sqrt{p}}
- \f{\lo^2 \Delta}{4 p} \bigg)
- \f{\ga^2 \lo^2 \sqrt\Delta}{6} \sqrt{|p|} \Bigg)
D_2(v) \, \Psi(v,\phi) \Bigg] ~,
\ea
where $D_2(v) = \Big| |v + 1| - |v - 1| \Big|$ and
\ba
C^+(v) &=& \f{|v + 2|}{4} \, \Big| |v + 1| - |v + 3| \Big| ~, \\
C^-(v) &=& C^+(v - 4), ~~~ \mathrm{and} ~~~ C^0(v) = - C^+(v) - C^-(v) ~.
\ea
Thus, the loop quantization of the $k=1$ FLRW model sourced with a massless scalar field
gives a quantum difference equation which couples states in intervals of four in the
volume variable $v$. The quantum constraint equation has the form of a Klein-Gordon
equation where $\phi$ plays the role of an emergent time and $\Theta$ is akin to a
time-independent spatial Laplacian operator.  As in the Klein-Gordon theory, the physical
Hilbert space can be constructed from the positive frequency solutions, which is
separable with sectors labeled by $\varepsilon$: $v \in \pm |\varepsilon| + 4 n$
where $n$ is an integer. These sectors are preserved by the dynamics, and hence there
is a super-selection in $v$. The physical states satisfy:
\be \label{k1ip}
\langle \Psi_1 | \Psi \rangle_\varepsilon = \sum_v B(v)
\bar \Psi_1(v,\phi_o) \Psi_2(v,\phi_o),
\ee
where the sum over $v$ refers to the choice of a particular super-selected sector $\varepsilon$.
Also, the physical states satisfy $\Psi(v,\phi) = \Psi(-v,\phi)$, i.e., they are symmetric
under the change of the orientation of the triads. Given that there are no fermions in the model,
this is a reasonable requirement on the physical states.  In order to extract physical predictions,
a family of Dirac obserables can be introduced: the field momentum $\hat p_\phi$ and the operator
$\hat v|_\phi$ which measures the volume with respect to the time $\phi$, as defined by
\be
|\hat v |_{\phi_o} \Psi(v,\phi) = e^{i \sqrt{\Theta} (\phi - \phi_o)} |v| \, \Psi(v,\phi_o) ~.
\ee
The Dirac observables are self-adjoint with respect to the inner product \eqref{k1ip}. The
expectation values of these Dirac observables can be computed by choosing suitable states,
for example Gaussian states peaked around classical trajectories at late times. Numerical
simulations of these states show the existence of a bounce, which can also be captured by
the effective dynamics obtained from an effective Hamiltonian corresponding to the quantum
constraint \eqref{k1qc} \cite{Ashtekar:2006es}.

\subsection{Comparison of Different Effective Descriptions of the $k=1$ Model}
\label{ss.comp}

The $k=1$ FLRW model provides important lessons about the effective dynamics in LQC.  In
this section we will examine the effective Hamiltonians for three different quantization
prescriptions of the closed FLRW space-time in LQC. The first prescription, reviewed in
Sec.\ \ref{ss.closed-standard}, is to express
the field strength of the connection in terms of holonomies around a closed loop of Planck
area.  The resulting effective Hamiltonian comes directly from the constraint \eqref{k1qc}
and we will call it the `F' effective description. The second one arises from starting from
the classical constraint \eqref{k1_class} and polymerizing the connection $A_a^i$. We refer
to this effective description as the `A' quantization. The third way to quantize the $k=1$
model is the `K' prescription where it is the extrinsic curvature is polymerized.

In the following comparison of the three effective theories, we will focus our attention on
the expansion scalar, which as discussed in the Introduction captures the detailed behaviour
of geodesics in the continuum space-time.  Our analysis will be based on the important
assumption that the effective Hamiltonians for all three prescriptions can be
consistently derived from a geometric formulation of quantum theory.

\subsubsection{Effective theory for the field strength `F' quantization}
\label{sss.cl-f}

The effective Hamiltonian corresponding to the holonomy quantization of the field strength
of the connection $A_a^i$ is given by
\be \label{effham1}
\mC_H^{\rm eff(F)} = - \f{3 \, p^{3/2}}{8 \pi G \gamma^2 \Delta} \bigg[
\sin^2\left(\bar \mu\left(c - \f{\lo}{2}\right)\right) - \sin^2 \sqrt\f{\Delta}{p}
+ (1 + \gamma^2) \f{\lo^2 \Delta}{4 p} \bigg]   + \f{p_\phi^2}{2 p^{3/2}} \approx 0 ~,
\ee
where without any loss of generality we have fixed the orientation of the triads to be positive,
and we work in the limit of $v \gg 1$ in which case inverse volume modifications can be ignored.
It is important to note that unlike the effective Hamiltonian for the $k=0$ model in LQC, the
above Hamiltonian cannot be obtained by the direct substitution of
$c \rightarrow \sin(\bar \mu c)/\bar \mu$ in the classical Hamiltonian constraint \eqref{k1_class}.
This is an important cautionary lesson that the polymerization of $c$ does not in general give
the correct effective Hamiltonian constraint.

We are interested in computing the properties of the expansion scalar $\theta$ of the geodesics
in the effective space-time, which is given by
\be
\theta = \f{\dot V}{V} = 3H = \f{3 \dot p}{2 p} ~.
\ee
As explained in the Introduction, the expansion scalar is a geometric quantity that
describes the evolution of geodesics.  To compute $\theta$,
we use the Hamilton's equation for $p$,
\be
\dot p = - \f{8 \pi G \gamma}{3} \cdot \f{\partial}{\partial c} \mC_H
= \f{2 p}{\gamma \sqrt\Delta} \sin\left(\bar \mu \left(c - \f{\lo}{2}\right)\right)
\cos\left(\bar \mu \left(c - \f{\lo}{2}\right)\right) ~,
\ee
and then the expansion scalar can be written as
\be \label{theta_fieldstrength}
\theta = \nonumber \f{3}{2 \gamma \sqrt{\Delta}}
\sin\left(2 \bar \mu \left(c - \f{\lo}{2}\right)\right) ~.
\ee
It is clear that, due to the underlying discreteness of the quantum geometry captured
in $\Delta$, the expansion scalar is bounded above:
\be
\theta_{\rm{max}} =  \f{3}{2 \gamma \sqrt{\Delta}}.
\ee
Also, note that the maximum value of expansion scalar is consistent with the maximum
value of the energy density that is typically observed in numerical simulations,
$\rho \approx 0.41 \rho_{\rm Pl}$.

\subsubsection{Effective theory for the `A' quantization}
\label{sss.cl-a}

The `A' connection quantization of the $k=1$ model has been performed in \cite{Corichi:2011pg}.
In Ref.\ \cite{Gupt:2011jh}, it was pointed out that the expansion scalar for the effective
Hamiltonian in this quantization is unbounded unless one considers inverse volume
modifications. Our argument here essentially follows that of Ref.\ \cite{Gupt:2011jh} except
that we focus on the role of holonomy modifications and ignore any possible inverse volume
corrections. The effective Hamiltonian for the connection quantization is obtained by
polymerizing $c$ in Eq.~\eqref{k1_class} by $c \rightarrow \sin(\bar \mu c)/\bar \mu$,
\be
\mC_H^{\rm eff(A)} = - \f{3 p^{3/2}}{8 \pi G \gamma^2} \bigg[\left(
\f{\sin(\bar \mu c)}{\sqrt\Delta} - \f{\lo}{2 \sqrt{p}} \right)^2
+ \f{\gamma^2 \lo^2}{4 p} \bigg] + \f{p_\phi^2}{2 p^{3/2}} \approx 0 ~.
\ee
Hamilton's equation for the triad then gives
\be
\dot p = \f{p}{\gamma} \bigg[\f{\sin(2 \bar \mu c)}{\sqrt{\Delta}}
- \f{\lo}{\sqrt{p}} \cos(\bar \mu c)\bigg],
\ee
which, using $\theta = 3 H = 3 \dot p/2 p$, results in the following expression for the
expansion scalar:
\be\label{theta_A}
\theta = \f{3}{2 \gamma \sqrt{\Delta}} \bigg[\sin(2 \bar \mu c)
- \f{\lo \sqrt\Delta}{p^{1/2}} \cos(\bar \mu c)\bigg].
\ee
Unlike the `F' field strength loop quantization, in the `A' quantization $\theta$
fails to have an upper bound. Due to the presence of the second term in the above
expression, $\theta$ can become arbitrarily large if dynamical trajectories approach
$p \to 0$.

Here it is important to note the following points. First,
within the underlying assumptions of the derivation of the effective framework, one can
safely conclude that \eqref{theta_A} has no maximum value in the effective space-time
description. Second, if one assumes that the effective spacetime description can be
consistently derived even for volumes smaller than the Planck volume, then one cannot
a priori rule out that for certain other types of matter, dynamical trajectories do
indeed reach the neighbourhood of $p=0$, which would cause the expansion scalar to
diverge.  Finally, even if the dynamical evolution avoids singularities, in this
quantization of $k=1$ FLRW model there is no general upper bound of the expansion scalar
where a bounce would necessarily occur.  This indicates that in this model, the bounce
may occur at drastically different values of the expansion scalar.  It is then clear
that the expansion found in the effective dynamics for the `A' loop quantization has
strikingly different properties from the expansion obtained in the standard loop
quantization of the $k=1$ FLRW model performed in \cite{Ashtekar:2006es}.

We conclude with the observation that the problem of the potential divergence of the expansion
scalar appears to be resolved if one considers inverse volume modifications in the effective
Hamiltonian for the `A' quantization \cite{Corichi:2013usa}, as suggested in \cite{Gupt:2011jh}.
However, even then the qualitative and quantitative features of the expansion scalar are
significantly different from the `F' quantization of \cite{Ashtekar:2006es}.  Thus,
the effective theory for the `A' loop quantization obtained by the polymerization of the
connection in the classical constraint does not by itself yield the standard upper bound on the
expansion scalar, which was a hallmark of the various bounce models in LQC where the quantization
is performed following the usual approach of expressing the field strength in terms of
holonomies of the connection.

\subsubsection{Effective theory for the `K' quantization}
\label{sss.cl-k}

The effective Hamiltonian for the `K' loop quantization of the $k=1$ FLRW model in
LQC is obtained in a similar manner as the one for the `A' quantization, however,
instead of polymering the connection in Eq.\ \eqref{k1_class}, we express the
constraint in terms of the extrinsic curvature which is then polymerized.  The
extrinsic curvature $K^i_a$ is given by
\be
K^i_a = \f{k}{\lo} \ow^i_a
\ee
where for $N=1$, we have $k = \lo \dot a = \dot p/(2 \sqrt{p})$.  Using Hamilton's
equation for $\dot p$ coming from the classical constraint \eqref{k1_class}, we
can related $k$ to connection $c$ by
\be \label{kc_rel}
\gamma k =  \left(c - \f{\lo}{2} \right) ~,
\ee
and it follows that the extrinsic curvature $k$ is conjugate to the triad $p$,
\be
\{k, p\} = \f{8 \pi G}{3} ~.
\ee
Then, by using \eqref{kc_rel}, \eqref{k1_class} can be rewritten as
\be
\mC_H^{\rm class} = - \f{3 p^{1/2}}{8 \pi G \gamma^2} \left(\gamma^2 k^2
+ \f{\lo^2 \gamma^2}{4} \right) + \f{p_\phi^2}{2 p^{3/2}} \approx 0 ~.
\ee
The last step is to polymerize the extrinsic curvature via
$\gamma k \rightarrow \sin(\bar \mu \gamma k)/\bar \mu$, giving the
effective Hamiltonian constraint
\be
\mC_H^{\rm eff(K)} = - \f{3 p^{1/2}}{8 \pi G \gamma^2} \bigg[\f{\sin^2(\bar \mu \gamma k)}{\bar \mu^2}
+ \f{\gamma^2 \lo^2}{4} \bigg] + \f{p_\phi^2}{2 p^{3/2}} \approx 0 ~.
\ee
Now Hamilton's equation for $p$ is
\be
\dot p = \f{p}{\gamma \sqrt\Delta} \, \sin(2 \bar \mu \gamma k),
\ee
which results in the following expression for the expansion scalar:
\ba
\theta &=& \f{3}{2 \gamma \sqrt{\Delta}} \, \sin(2 \bar \mu \gamma k) \nn \\
&\leq& \f{3}{2 \gamma \sqrt{\Delta}} ~.
\ea
Thus, in striking contrast to the `A' quantization, the expansion scalar in the `K' quantization
is not only generically bounded, but in addition its maximal value is exactly the same as the one
found in the `F' loop quantization.  Hence, the `K' quantization
for some reason captures extremely well the qualitative details of the bounce regime of the
original quantization of the $k=1$ FLRW model. It is important to note that unlike the case of
the `A' quantization, there is no need to incorporate inverse volume corrections in order to
generically resolve geodesic singularities in the effective theory.

To summarize, we find that the effective dynamics of the `K' quantization of the $k=1$ FLRW model
are surprisingly similar to the effective dynamics of the standard `F' loop quantization,
and in fact also similar to all of the other isotropic models quantized in LQC.  This is in
contrast to the `A' loop quantization where the qualitative features of the effective dynamics
are significantly different to the other LQC models.  This indicates that the `K' loop
quantization is a more harmonious choice that captures the main features of the standard
LQC quantization of isotropic cosmologies.  In the following section, we shall show that
this conclusion also holds true for the Bianchi IX model.

\section{The Bianchi IX Space-time}
\label{s.b-ix}

In the previous section, we saw how the polymerization of the
extrinsic curvature captures the main qualitative features of the
full loop quantization of the closed FLRW space-time.  On the other
hand, the polymerization of the Ashtekar-Barbero connection ---the
approach used in the loop quantization of the Bianchi II and Bianchi
IX space-times in \cite{Ashtekar:2009um, WilsonEwing:2010rh}--- fails
to capture some of the more important features of the full loop
quantization of the closed FLRW universe.

This motivates developing the `K' loop quantization of the Bianchi
IX model, which we shall perform in this section.  (The `K' loop
quantization of the Bianchi II model is given in Appendix \ref{s.b-ii}.)
In addition, we shall compare the effective theories for the `A' and `K'
loop quantizations, and we shall see that the `K' prescription has many
of the same properties as the LQC of the Bianchi I model, while the `A'
effective theory does not.  This suggests that the `K' loop quantization
does a better job of capturing the LQC effects in space-times that are
spatially curved.

We begin this section by reviewing the classical Hamiltonian framework for
the Bianchi IX space-time, then study the effective theory of the `A' loop
quantization before presenting the `K' loop quantization of Bianchi IX.  We
end by showing that in the `K' effective theory, the expansion and shear
are bounded by the same upper limits as in the effective theory of the LQC
of Bianchi I.

\subsection{A Brief Review of the Bianchi IX Cosmology}
\label{ss.rev-b-ix}

The Bianchi IX space-time is a generalization of the closed FLRW cosmology,
in that the fiducial triads are again given by \eqref{tri-1}--\eqref{tri-3}
and the fiducial co-triads by \eqref{co-1}--\eqref{co-3}, with the difference
that there are now three independent scale factors $a_i(t)$,
\be
\om_a^i = a^i(t) \, \ow_a^i,
\ee
where it is understood that there is no sum over repeated internal indices
that are both contravariant or covariant, as is the case for $i$ here.

In terms of LQC variables, this becomes
\be
E^a_i = \f{p_i}{\lo^2} \, \sqrt{\oq} \, \oe^a_i, \qquad
A_a^i = \f{c^i}{\lo} \, \ow_a^i,
\ee
and the $p_i$ are related to the scale factors by, e.g., $p_1 = a_2 a_3 \lo^2$.
Again, there is no sum over the index $i$ in these two relations.

The spin-connection has a slightly more complicated form in the anistropic
setting, for example
\be
\Gamma_a^1 = \f{1}{2} \left( \f{p_2 p_3}{p_1} - \f{p_2}{p_3}
- \f{p_3}{p_2} \right) \ow_a^1;
\ee
the two other spin-connections can be obtained by cyclic permutations.

Finally, the non-zero Poisson brackets are given by
\be
\{c_i, p_j\} = 8 \pi G \ga \, \de_{ij}.
\ee

Just as in the closed FLRW model, the Gauss and diffeomorphism constraints
are automatically satisfied due to the form of $E^a_i$ and $A_a^i$, which
leaves only the scalar constraint, which for $N=1$ is
\begin{align} \label{h-ix-c}
\mC_H = & \, - \f{1}{8 \pi G \ga^2 \sqrt{p_1 p_2 p_3}} \bigg[ p_1 p_2 c_1 c_2
+ p_2 p_3 c_2 c_3 + p_3 p_1 c_3 c_1
+ \lo \big( p_1 p_2 c_3 + p_2 p_3 c_1 + p_3 p_1 c_2 \big) \nn \\
& + \f{\lo^2 (1 + \ga^2)}{4} \bigg( 2 p_1^2 + 2 p_2^2 + 2 p_3^2
- \f{p_1^2 p_2^2}{p_3^2} - \f{p_2^2 p_3^2}{p_1^2}
- \f{p_3^2 p_1^2}{p_2^2} \bigg) \bigg]
+ \f{p_\phi^2}{2 \sqrt{p_1 p_2 p_3}} \approx 0.
\end{align}
The scalar constraint determines the dynamics and, combined with the Poisson
brackets, this is sufficient to study the classical dynamics of the Bianchi
IX space-time.

From the constraint, the equations of motion can be obtained, for example
\be
\d p_1 = \f{1}{\ga} \left( \sqrt\f{p_1 p_2}{p_3} \, c_2 +
\sqrt\f{p_1 p_3}{p_2} \, c_3 + \lo \sqrt\f{p_2 p_3}{p_1} \right),
\ee
while $\d p_2$ and $\d p_3$ can be obtained by cyclic permutations.
Then, the directional Hubble rate $H_1$ is given by
\begin{align}
H_1 &= \f{\d a_1}{a_1} = \f{\d p_2}{2 p_2} + \f{\d p_3}{2 p_3}
- \f{\d p_1}{2 p_1} \nn \\ &
= \f{1}{\ga} \left[ \sqrt\f{p_1}{p_2 p_3} \, c_1
+ \f{\lo}{2} \left( \sqrt\f{p_1 p_2}{p_3^3} + \sqrt\f{p_1 p_3}{p_2^3}
- \sqrt\f{p_2 p_3}{p_1^3} \right) \right].
\end{align}
The quantities of interest in the singularity theorems of Penrose,
Hawking and Geroch are the expansion $\theta$ and the shear $\sigma_{ab}$,
which in the Bianchi space-times are related to the directional Hubble
rates as follows:
\begin{align}
\theta = & \: H_1 + H_2 + H_3, \\
\sigma^2 = & \sigma_{ab} \sigma^{ab} = \f{1}{3} \Big[ (H_1 - H_2)^2
+ (H_2 - H_3)^2 + (H_3 - H_1)^2 \Big].
\end{align}
If the quantities of $\theta$ and $\sigma^2$ can be shown to be bounded,
this indicates that geodesic singularities may be resolved.  Of course, in
the classical theory neither $\theta$ nor $\sigma^2$ are bounded in the
Bianchi IX space-time, and so singularities do exist, most notably the big
bang and the big crunch.

Before ending this section, it is useful to rewrite the scalar constraint
in terms of different variables.  As we saw in the previous section, in
some circumstances it can be useful to work with the extrinsic curvature
$K_a^i$ rather than $A_a^i$.  Therefore, introducing
\be
K_a^i = \f{k^i}{\lo} \, \ow_a^i,
\ee
where there is no sum over $i$, it follows from the definition of
$A_a^i = \Gamma_a^i + \ga K_a^i$ and the fact that $\Gamma_a^i$ is
a function of $p_i$ only that
\be
\{k_i, p_j\} = 8 \pi G \, \de_{ij}.
\ee
From the form of $\Gamma_a^i$, it is possible to explicitly check that
$\{k_i, k_j\} = 0$, as expected.

The Hamiltonian constraint in terms of $p_i$ and $k_i$ is now
\begin{align} \label{h-k}
\mC_H = & \, - \f{1}{8 \pi G \sqrt{p_1 p_2 p_3}} \bigg[ p_1 p_2 k_1 k_2
+ p_2 p_3 k_2 k_3 + p_3 p_1 k_3 k_1 \nn \\ &
+ \f{\lo^2}{4} \bigg( 2 p_1^2 + 2 p_2^2 + 2 p_3^2 - \f{p_1^2 p_2^2}{p_3^2}
- \f{p_2^2 p_3^2}{p_1^2} - \f{p_3^2 p_1^2}{p_2^2} \bigg) \bigg]
+ \f{p_\phi^2}{2 \sqrt{p_1 p_2 p_3}} \approx 0.
\end{align}
Of course, classically it is completely equivalent to work with either of the
variables $c_i$ or $k_i$.  The important question here is whether one choice
is more convenient to pass to the quantum theory.  It is immediately apparent
that the constraint \eqref{h-k} is somewhat simpler, but in loop quantum
gravity, it is holonomies of the connection $A_a^i$ which are viewed as
fundamental operators rather than functions of the extrinsic curvature, which
seems to indicate that we should work with \eqref{h-ix-c}.

Despite this point, in the previous section we saw how a polymerization of $K_a^i$
---but not the polymerization of $A_a^i$--- captures the most important qualitative
characteristics of the closed FLRW model in LQC.  Therefore, it is important
to study the qualitative physics of a `K' polymerization of the Bianchi
IX space-time, which is after all a generalization of the closed Friedmann
universe.  It will also be interesting to compare the effective theories of
the `A' and `K' loop quantizations, so we will first study some of the
properties of the `A' effective theory, before deriving the `K' loop
quantization of the Bianchi IX model, and then studying its effective
theory as well.

\subsection{Effective Theory of the `A' Quantization}
\label{ss.ab}

As explained in the Introduction, it is not possible to quantize the Bianchi IX
model in LQC by expressing the field strength in terms of the holonomies of
$A_a^i$ since the resulting expression is not almost periodic in $c_i$.
Because of this difficulty, only the `A' and `K' loop quantizations are
possible for Bianchi IX.  In this part, we will review some of the properties
of the effective theory of the `A' quantization; for details regarding the
full quantum theory, see \cite{WilsonEwing:2010rh}.

The effective scalar constraint (ignoring inverse volume corrections) for the
`A' loop quantization of Bianchi IX is \cite{WilsonEwing:2010rh}
\begin{align}
\mC_H = & \, - \f{\sqrt{p_1 p_2 p_3}}{8 \pi G \ga^2} \bigg[
\f{1}{\De} \Big( \sin \b\mu_1 c_1 \sin \b\mu_2 c_2 + \sin \b\mu_2 c_2
\sin \b\mu_3 c_3 + \sin \b\mu_3 c_3 \sin \b\mu_1 c_1 \Big) \nn \\ &
+ \f{\lo}{\sqrt\De} \bigg( \f{p_1 p_2}{p_3} \sin \b\mu_3 c_3
+ \f{p_2 p_3}{p_1} \sin \b\mu_1 c_1 + \f{p_3 p_1}{p_2} \sin\b\mu_2 c_2 \bigg)
\nn \\ &
+ \f{\lo^2 (1 + \ga^2)}{4 p_1 p_2 p_3} \bigg( 2 p_1^2 + 2 p_2^2 + 2 p_3^2
- \f{p_1^2 p_2^2}{p_3^2} - \f{p_2^2 p_3^2}{p_1^2}
- \f{p_3^2 p_1^2}{p_2^2} \bigg) \bigg]
+ \f{p_\phi^2}{2 \sqrt{p_1 p_2 p_3}} \approx 0,
\end{align}
where
\be
\b\mu_1 = \sqrt \f{p_1 \Delta}{p_2 p_3}, \qquad
\b\mu_2 = \sqrt \f{p_2 \Delta}{p_3 p_1}, \qquad
\b\mu_3 = \sqrt \f{p_3 \Delta}{p_1 p_2}.
\ee

It follows that
\be
\d p_1 = \ga^{-1} \left[ \f{p_1}{\sqrt\De} \left( \sin \b\mu_2 c_2
+ \sin \b\mu_3 c_3 \right) + \lo \sqrt \f{p_2 p_3}{p_1} \right]
\cos \b\mu_1 c_1,
\ee
and so the directional Hubble rate $H_1$ is
\begin{align}
H_1 = & \, \f{1}{2 \ga \sqrt\De} \Big( \sin (\b\mu_1 c_1 - \b\mu_2 c_2)
+ \sin (\b\mu_1 c_1 - \b\mu_3 c_3) + \sin (\b\mu_2 c_2 + \b\mu_3 c_3)
\Big) \nn \\ &
+ \f{\lo}{2 \ga} \left( \sqrt\f{p_1 p_3}{p_2^3} \cos \b\mu_2 c_2
+ \sqrt\f{p_1 p_2}{p_3^3} \cos \b\mu_3 c_3 - \sqrt\f{p_2 p_3}{p_1^3}
\cos \b\mu_1 c_1 \right),
\end{align}
the other directional Hubble rates can be obtained by cyclic
permutations.  The directional Hubble rates determine the expansion,
\begin{align}
\theta = & \, \f{1}{2 \ga \sqrt\De} \Big( \sin (\b\mu_1 c_1 + \b\mu_2 c_2)
+ \sin (\b\mu_2 c_2 + \b\mu_3 c_3) + \sin (\b\mu_3 c_3 + \b\mu_1 c_1)
\Big) \nn \\ &
+ \f{\lo}{2 \ga} \left( \sqrt\f{p_2 p_3}{p_1^3} \cos \b\mu_1 c_1
+ \sqrt\f{p_1 p_3}{p_2^3} \cos \b\mu_2 c_2
+ \sqrt\f{p_1 p_2}{p_3^3} \cos \b\mu_3 c_3 \right),
\end{align}
and it is easy to see that while the terms on the first line are bounded, this
is not the case for the terms on the second line.  As there is no clear bound
on $\theta$, at this point it is impossible to determine whether singularities
are resolved in the full theory without performing a significantly more
detailed analysis.  An important point is that by including inverse volume
corrections, it is possible to show that $\theta$ is bounded \cite{Gupt:2011jh},
but the resulting bound is much weaker than the one obtained for the LQC of
the Friedmann and Bianchi I space-times.

While we have not given the explicit form of $\sigma^2$ here, it is qualitatively
very similar to $\theta$ in that it is necessary to include inverse volume
corrections in order to show that it is bounded \cite{Gupt:2011jh}, and once
again the resulting bound is quite weak compared to the limit on $\sigma^2$ found
in the Bianchi I space-time.

\subsection{The `K' Loop Quantization: Kinematical Hilbert Space}
\label{ss.k-b-ix-kin}

Since the `A' loop quantization of the Bianchi IX model was studied in
\cite{WilsonEwing:2010rh}, we simply quoted the resulting effective
equations without reviewing the entire quantum theory.  On the other hand,
the `K' loop quantization of the Bianchi IX model has not yet been developed
using $\b\mu_i$ holonomies, although a $\mu_o$-type%
\footnote{The $\mu_o$ quantization scheme was historically
the first to be developed in LQC, where instead of Eq.\
\eqref{bar-mu}, one sets $\mu_o = {\rm constant}$.  However,
the $\mu_o$ framework has several pathologies, including
an incorrect semi-classic limit and an unphysical dependence
on the choice of the fiducial cell \cite{Ashtekar:2006wn}
and so it is no longer considered a viable quantization
scheme.  These pathologies do not appear in the ``improved
dynamics'' of $\b\mu$ LQC models where one sets the length
of the fundamental holonomy operators to $\b\mu$ [see
Eq.\ \eqref{bar-mu}].}
`K' quantization was given in \cite{Bojowald:2003xf}.  Therefore, here
we will present the $\b\mu_i$ `K' loop quantization, starting by
defining the kinematical Hilbert space and then deriving the Hamiltonian
constraint operator in the next section.  While the kinematical Hilbert
space is essentially the same as the one given in \cite{Bojowald:2003xf,
WilsonEwing:2010rh}, the Hamiltonian constraint operator is
significantly different.

As is usual, the basis for the gravitational sector in loop quantum cosmology
is given by the Bohr compactification of the real line.  Explicitly, the
basis states are eigenvectors of the $\h p_i$ operator,
\be \label{basis}
\h p_i | p_1, p_2, p_3 \ket = p_i | p_1, p_2, p_3 \ket,
\ee
and the inner product between two such states is given by the Kronecker
delta,
\be
\bra p_1, p_2, p_3 | q_1, q_2, q_3 \ket = \de_{p_1, q_1} \de_{p_2, q_2}
\de_{p_3, q_3}.
\ee
A generic state is given by a countable sum of the form
\be
\Psi = \sum_{p_1, p_2, p_3} \alpha_{p_1, p_2, p_3} | p_1, p_2, p_3 \ket,
\ee
and the wave function $\Psi$ is normalized if
\be
\sum_{p_1, p_2, p_3} |\alpha_{p_1, p_2, p_3}|^2 = 1.
\ee

The other fundamental operator defined in the kinematical Hilbert space of the
`K' loop quantization are complex exponentials of the extrinsic curvature,
\be
\wh {e^{-i \mu k_1}} |p_1, p_2, p_3 \ket = |p_1 + 8 \pi \lp^2 \mu, p_2, p_3 \ket.
\ee
The action of operators corresponding to complex exponentials of $k_2$ and $k_3$
is in direct analogy with that of $k_1$.

\subsection{The `K' Loop Quantization: The Hamiltonian Constraint Operator}
\label{ss.k-b-ix-ch}

In order to define the Hamiltonian constraint operator, it is necessary to (i) express
all $k_i$ terms in the classical Hamiltonian as complex exponentials, and (ii) inverse
triad operators must also be defined.

Let us start by recalling the procedure in the `A' loop quantization of the Bianchi
IX space-time, where the non-local operator corresponding to the connection $A_a^j$ is
obtained by taking the holonomy of length $\b\mu_j$ parallel to $\oe^a_j$,
(see \cite{WilsonEwing:2010rh} for details)
\be \label{def-a}
\wh{A_a} = \sum_j \f{1}{2 \b\mu_j \lo} \left[
e^{2 \b\mu_j c_j \tau^j} - e^{- 2 \b\mu_j c_j \tau^j} \right], \qquad
\Rightarrow \quad
\wh{c_j} = \f{1}{2 i \b\mu_j} \left[ e^{i \b\mu_j c_j} - e^{-i \b\mu_j c_j} \right],
\ee
where
\be
\b\mu_1 = \sqrt \f{p_1 \Delta}{p_2 p_3}, \qquad
\b\mu_2 = \sqrt \f{p_2 \Delta}{p_1 p_3}, \qquad
\b\mu_3 = \sqrt \f{p_3 \Delta}{p_1 p_2},
\ee
and $A_a = A_a^j \tau_j$, recall that the $\tau_j$ are a basis of $\mathfrak{su}(2)$.
Note that in the two equations \eqref{def-a}, the index $j$ is not summed over inside
the argument of any of the exponentials; the only sum over $j$ is shown explicitly.

In the `K' loop quantization, we follow the same procedure for the extrinsic curvature
rescaled by the Immirzi parameter $\ga K_a^i$ (rather than $A_a^i$ as is done in the `A'
quantization), and obtain the analogous result
\be \label{k-hat}
\wh{k_j} = \f{1}{2 i \b\mu_j \ga} \left[ e^{i \b\mu_j \ga k_j}
- e^{-i \b\mu_j \ga k_j} \right]
= \wh{ \f{\sin \b\mu_j \ga k_j}{\b\mu_j \ga} }.
\ee

Since the length $\b\mu_j$ depends on the phase space variables $p_i$, the action of
$\sin \b\mu_j \ga k_j$ on a state in the basis \eqref{basis} is not trivial.  It is
convenient to introduce the variables
\be
\la_i = \f{\sqrt p_i}{(4 \pi G \hbar \ga \sqrt \Delta)^{1/3}},
\qquad v = 2 \la_1 \la_2 \la_3 = \f{V}{2 \pi G \hbar \ga \sqrt \Delta}.
\ee
(Note that this is the same $v$ as appears in the LQC of the closed model that was
presented in Sec.\ \ref{ss.closed-standard}.)  Then,
\begin{align}
\sin \b\mu_1 \ga k_1 |\la_1, \la_2, \la_3\ket &
= \f{1}{2i} \Big[ |\la_1 - (\la_2 \la_3)^{-1}, \la_2, \la_3\ket
- |\la_1 + (\la_2 \la_3)^{-1}, \la_2, \la_3\ket \Big] \nn \\ &
= \f{1}{2i} \bigg[ |\f{v-2}{v} \cdot \la_1, \la_2, v-2\ket
- |\f{v+2}{v} \cdot \la_1, \la_2, v+2\ket \bigg],
\end{align}
where on the second line we use the basis $(\la_1, \la_2, v)$, which
is viable for states that do not have any support on $v = 0$.

The next step is to define inverse triad operators.  There are many ambiguities
that must be dealt with at this point, but since we shall study inverse triad
operators in some detail in Sec.\ \ref{s.inv}, we shall choose the
simplest inverse triad operator possible here.  So, for the sake of simplicity,
we take
\be \label{inv-v-app}
\wh {\, \f{1}{\la_i} \,} \: | \la_1, \la_2, \la_3 \ket = \begin{cases}
0 & \text{if  } \: \wh{\la_i} | \la_1, \la_2, \la_3 \ket = 0,\\
\la_i^{-1} \: | \la_1, \la_2, \la_3 \ket \:\:& \text{otherwise.}\\
\end{cases}
\ee
It should be kept in mind that many other choices of inverse triad operators
are possible, as we shall see in Sec.\ \ref{s.inv}.  However, this
freedom is not important for the analysis of this section.

Choosing a symmetric factor ordering, the Hamiltonian constraint
operator for the lapse $N=1$ is
\begin{align}
\wh{\mC_H} = & \, - \f{1}{8 \pi G} \bigg[ \f{\sqrt{V}}{2\ga^2 \Delta}
\Big( \sin \b\mu_1 \ga k_1 \sin \b\mu_2 \ga k_2
+ \sin \b\mu_2 \ga k_2 \sin \b\mu_1 \ga k_1
+ \sin \b\mu_2 \ga k_2 \sin \b\mu_3 \ga k_3 \nn \\ & \qquad
+ \sin \b\mu_3 \ga k_3 \sin \b\mu_2 \ga k_2
+ \sin \b\mu_1 \ga k_1 \sin \b\mu_3 \ga k_3
+ \sin \b\mu_3 \ga k_3 \sin \b\mu_1 \ga k_1 \Big) \sqrt{V} \nn \\ &
+ \f{\lo^2}{4 V} \bigg( 2 p_1^2 + 2 p_2^2 + 2 p_3^2
- \f{p_1^2 p_2^2}{p_3^2} - \f{p_2^2 p_3^2}{p_1^2}
- \f{p_3^2 p_1^2}{p_2^2} \bigg) \bigg] + \f{p_\phi^2}{2 V},
\end{align}
where we have dropped the hats on the righthand side in order to
simplify the notation.  Different factor orderings are of course
possible, but seem to change little in the resulting physics so long
as the physical size of the universe remains much larger than the
Planck volume at all times, at least in the simplest models where
different factor orderings have been studied in some detail
\cite{MenaMarugan:2011me}.

As an aside, we point out that we have assumed that the $p_i$ are always
positive.  It is also possible to allow negative $p_i$ as well (this
corresponds to allowing different orientations of the triads), in which
case the Hamiltonian is the same, modulo some additional $\sgn(p_i)$
terms that appear (see e.g.\ \cite{MartinBenito:2009aj,
WilsonEwing:2010rh} for details in different settings, it is easy to
adapt the calculations to the Bianchi IX space-time).  However, the
qualitative behaviour of the Hamiltonian constraint operator is not
significantly altered by the $\sgn(p_i)$ terms, and therefore we
shall ignore them here.

It is clear that singular states (i.e., states where $\h V |\Psi\ket = 0$)
are automatically annihilated by the Hamiltonian constraint operator due
to the properties of the inverse triad operator.  For non-singular states
(i.e., states that have no support on any singular state), the quantum
constraint equation $\wh{\mC_H} \Psi = 0$ implies
\begin{align}
\hbar^2 \partial_\phi^2 \Psi = \, &
\f{\hbar V}{16 \ga \sqrt\Delta} \Big[
\sqrt{v(v+4)} \Psi^+_4 - v \Psi^+_0
- v \Psi^-_0 + \sqrt{v(v-4)} \Psi^-_4 \Big] \nn \\ &
- \f{\lo^2}{32 \pi G} \bigg( 2 p_1^2 + 2 p_2^2 + 2 p_3^2
- \f{p_1^2 p_2^2}{p_3^2} - \f{p_2^2 p_3^2}{p_1^2}
- \f{p_3^2 p_1^2}{p_2^2} \bigg) \Psi,
\end{align}
where it is understood that $\Psi := \Psi(\la_1, \la_2, v; \phi)$,
the operator corresponding to $p_\phi$ is given by
$\wh{p_\phi} = -i\hbar \partial_\phi$, and
\begin{align} \label{psi-pm}
\Psi^\pm_m = \, &
\Psi\left(\f{v \pm m}{v \pm 2} \cdot \la_1,
\f{v \pm 2}{v} \cdot \la_2, v \pm 4; \phi\right)
+ \Psi\left(\f{v \pm m}{v \pm 2} \cdot \la_1,
\la_2, v \pm 4; \phi\right) \nn \\ &
+ \Psi\left(\f{v \pm 2}{v} \cdot \la_1,
\f{v \pm m}{v \pm 2} \cdot \la_2, v \pm 4; \phi\right)
+ \Psi\left(\f{v \pm 2}{v} \cdot \la_1,
\la_2, v \pm 4; \phi\right) \nn \\ &
+ \Psi\left(\la_1,
\f{v \pm m}{v \pm 2} \cdot \la_2, v \pm 4; \phi\right)
+ \Psi\left(\la_1,
\f{v \pm 2}{v} \cdot \la_2, v \pm 4; \phi\right).
\end{align}
Just as was seen in Sec.\ \ref{ss.closed-standard} for the closed FLRW
model, this quantum constraint equation can be used in order to study the
full quantum dynamics of the `K' loop quantization of the Bianchi IX
space-time.  This equation has two particularly nice properties.
The first is that not only do the matter and gravitational sectors
decouple, but in addition, the scalar field $\phi$ can be used as
a relational clock variable.  Therefore, in this setting the problem
of time can be addressed by using a relational framework.

Another nice property of the Hamiltonian constraint operator is
that it resolves the big-bang and big-crunch singularities that
appear in the classical theory.  This can be seen in the following
manner.  First, the big-bang and big-crunch singularities correspond
to singular states, i.e., that have support on $v = 0$.  Then, a
state which is initially non-singular (i.e., one that has no support
on singular states) can be seen to remain non-singular at all times.
This is due to the fact that, using $\phi$ as a time variable, the
support of the wave function on $v$ can only evolve via the $\Psi_4^\pm$
terms in the quantum constraint equation.  But in order to gain
support on the point $v = 0$, we must have at some point $v - 4 = 0$.
However, in that case the prefactor to $\Psi_4^-$ will vanish, and
then the portion of the wave function which would have become
singular is annihilated and so the wave function remains non-singular.

Therefore, a non-singular state will never have any support on $v=0$, and
it is in this sense that the classical big-bang and big-crunch singularities
are resolved in the quantum theory.  As we shall see in the next part,
the singularity is resolved in an even stronger fashion in the effective
theory where the expansion and shear are bounded.

\subsection{Effective Theory of the `K' Quantization}
\label{ss.ext}

Now that the `K' loop quantization of the Bianchi IX model has been
constructed, it is possible to analyze its effective theory in order
to determine whether it automatically bounds geometric quantities
like the expansion and shear.

The Hamiltonian constraint for the effective theory of the `K' loop
quantization is
\begin{align}
\mC_H = & \, - \f{\sqrt{p_1 p_2 p_3}}{8 \pi G \ga^2} \bigg[
\f{1}{\De} \Big( \sin \b\mu_1 \ga k_1 \sin \b\mu_2 \ga k_2
+ \sin \b\mu_2 \ga k_2 \sin \b\mu_3 \ga k_3 + \sin \b\mu_3 \ga k_3
\sin \b\mu_1 \ga k_1 \Big) \nn \\ &
+ \f{\lo^2}{4 p_1 p_2 p_3} \bigg( 2 p_1^2 + 2 p_2^2 + 2 p_3^2
- \f{p_1^2 p_2^2}{p_3^2} - \f{p_2^2 p_3^2}{p_1^2}
- \f{p_3^2 p_1^2}{p_2^2} \bigg) \bigg]
+ \f{p_\phi^2}{2 \sqrt{p_1 p_2 p_3}} \approx 0,
\end{align}
from which it follows that
\be \label{p1-dot}
\d p_1 = \f{p_1}{\ga \sqrt\De} \left( \sin \b\mu_2 \ga k_2
+ \sin \b\mu_3 \ga k_3 \right) \cos \b\mu_1 \ga k_1.
\ee
From this equation of motion, it is easy to calculate the directional
Hubble rates, for example $H_1$ is
\be
H_1 = \f{1}{2 \ga \sqrt\De} \Big( \sin (\b\mu_1 \ga k_1 - \b\mu_2 \ga k_2)
+ \sin (\b\mu_1 \ga k_1 - \b\mu_3 \ga k_3) + \sin (\b\mu_2 \ga k_2 + \b\mu_3 \ga k_3)
\Big),
\ee
and the other directional Hubble rates $H_2$ and $H_3$ are obtained
via cyclic permutations.

It immediately follows that $\theta$ and $\sigma^2$ are bounded,
(note that there is no need to impose any energy conditions on the
matter fields)
\begin{align}
\theta & \, = \f{1}{2 \ga \sqrt\De} \Big( \sin (\b\mu_1 \ga k_1 + \b\mu_2 \ga k_2)
+ \sin (\b\mu_2 \ga k_2 + \b\mu_3 \ga k_3) + \sin (\b\mu_3 \ga k_3 + \b\mu_1 \ga k_1) \Big)
\nn \\ & \le \f{3}{2 \ga \sqrt\De},
\end{align}
\begin{align}
\sigma^2 & = \f{1}{3 \ga^2 \De} \big[ \sin (\b\mu_1 \ga k_1 - \b\mu_2 \ga k_2)
- (\cos \b\mu_1 \ga k_1 - \cos \b\mu_2 \ga k_2) \sin \b\mu_3 \ga k_3 \big]^2
+ {\rm cyclic ~ permutations} \nn \\ &
\le \f{10.125}{3 \ga^2 \De},
\end{align}
where the bound on $\sigma^2$ is saturated when one of the
$\b\mu_i \ga k_i = \pi/6$, another is given by $\pi/2$ and the
last is $5 \pi / 6$.  Note that these bounds are exactly the
same as those obtained for the effective theory of the Bianchi
I model \cite{Corichi:2009pp, Gupt:2011jh}.  Following the
arguments given in the Introduction, these bounds are a strong
indication of the absence of geodesic singularities in this
effective theory.

Thus, we see a major difference between the effective theories of
the `A' and `K' quantizations of the Bianchi IX model: in the `A'
quantization, the expansion and the shear are not bounded, while
in the `K' quantization both quantities are not only bounded, but
in addition the upper bounds are in perfect agreement with those of
the Bianchi I model.  This shows that there is a nice continuity
between the loop quantizations of the Bianchi I and Bianchi IX
space-times, at least for the `K' loop quantization.

\section{Inverse Triad Effects}
\label{s.inv}

As discussed in the Introduction, in canonical LQG there are two particularly
important steps that are necessary in order to pass from the classical to the
quantum theory, namely (i) expressing all occurrences of the Ashtekar-Barbero
connection (or its field strength) in terms of holonomies, and (ii) replacing
all inverse powers of a densitized triad (or its determinant) by appropriate
Poisson brackets that are equivalent to the original expression, but are devoid
of any inverse powers of the densitized triads \cite{Thiemann:1996aw}.
In LQC, these two steps are adapted to the symmetry-reduced context and in the
effective theory give, respectively, holonomy and inverse triad corrections to
the classical equations of motion.  So far in this paper, we have concentrated
on holonomy corrections, and in this section we move our focus to study inverse
triad corrections in spatially compact space-times in LQC.  We will work almost
entirely at the level of the effective theory in this section, and so care must
be taken when interpreting these results as the effective theory is valid only
if the underlying assumptions in its derivation are satisfied.

We will begin by describing the many ambiguities that arise in the definition of
inverse triad operators.  There are already many ambiguities for inverse volume
operators in the isotropic setting, and we would at first expect there to be
even more ambiguities present when anisotropies are present.  However, as we shall
see, if one uses the $\b\mu_i$ holonomies that naturally arise in LQC in order
to define the inverse triad operators, then one is lead to rewrite generic inverse
triad operators in terms of the basic inverse volume operator in which case there
are no new ambiguities.  We also propose an alternative inverse triad operator
which is free of ambiguities.

We end this section by considering the possibility that inverse triad corrections
may bound the matter energy density $\rho$.  In the previous section we saw how
in the effective theory for the `K' loop quantization, the expansion and shear
are bounded by holonomy corrections.  As we shall see, this is not the case for
the energy density.  However, it was recently shown that in the closed FLRW model
inverse triad corrections combined with holonomy corrections can in fact bound
the energy density \cite{Corichi:2013usa}.  Although this result gives hope that
the combination of holonomy and inverse triad corrections may bound the energy
density in the Bianchi IX model as well, we show that when anisotropies are
large, the energy density is not bounded on the constraint surface.

\subsection{Ambiguities in the Inverse Volume Operator}
\label{ss.amb-inv}

We open this part with a brief general discussion regarding the freedom of grouping
together (or not) different classical quantities when defining operators and
how this leads to ambiguities in the resulting quantum theory.  We then consider
ambiguities that enter into the definition of the basic inverse volume operator
corresponding to $1/V$, and show that some of these ambiguities are resolved if the
holonomies entering in the definition of the inverse volume operator are of
the $\b\mu_i$ type.  Throughout this section, we will work with the `K' loop
quantization of Bianchi IX, it is easy to extend the discussion to other
space-times or quantization prescriptions.

One of the first steps in the loop quantization procedure is to determine
where inverse triad operators should appear in the Hamiltonian constraint
operator.  Starting with the gravitational sector (we comment on the
matter sector later), in the Bianchi IX model inverse triad terms only
appear in the second term of the scalar constraint, which is of the
form
\be
\int N \sqrt q \: {}^{(3)}\!R.
\ee
LQC is a symmetry-reduced theory, and so these quantities are first expressed
in terms of the gauge-fixed densitized triad $E^a_i \sim p_i \oe^a_i$, and inverse
triad operators are introduced after this substitution.  Thus, it is important
to determine whether inverse triad operators should appear for the combination
of $N \sqrt q \: {}^{(3)}\!R$, or $\sqrt q \: {}^{(3)}\!R$, or perhaps
${}^{(3)}\!R$ alone.  In short, this choice can be summed up as
\be \label{options}
\wh{N \sqrt q \: {}^{(3)}\!R} \neq \wh{N}  \wh{\sqrt q \: {}^{(3)}\!R}
\neq \wh{N}  \wh{\sqrt q} \: \wh{{}^{(3)}\!R},
\ee
where these choices are not equivalent since, for most definitions of the inverse
triad operator,
\be
\wh{p_i^n} \wh{p_i^{-m}} \neq \wh{p_i^{n-m}}.
\ee

If the lapse is taken to be $N=1$, as is the case in this paper, then the
first two possibilities in \eqref{options} are equivalent, but this is
clearly not the case in general.  Precisely due to the fact that the lapse,
being a Lagrange multiplier (which can depend on phase space variables),
can be chosen freely, it has been argued that inverse triad effects should
not depend on the choice of a Lagrange multiplier in which case the first
option presented in \eqref{options} would not be a good choice
\cite{MartinBenito:2009aj}.  This is also related to the fact that it is
possible to remove all inverse triad operators by the choice of a specific
lapse \cite{Ashtekar:2007em}, a procedure which appears to lead to inequivalent
quantum theories in some settings \cite{Kaminski:2008td} (although not in
the flat FLRW model where this was first noticed in \cite{Ashtekar:2007em}).

Now, from a covariant perspective $\sqrt{-g} = N \sqrt q$, so it seems
unnatural to treat $N$ and $\sqrt q$ differently.  These two arguments
suggest that inverse triad operators should be defined based on the form
of ${}^{(3)}\!R$ alone, and this is what we will do in this paper.

However, it should be clear from the heuristic nature of the arguments
presented here that there is not yet a solid answer to this question, and input
from full LQG will be necessary in order to resolve this ambiguity.  At
this point, it is important to keep in mind that a different choice may
prove to be the correct one in the future.

The argument given above for the gravitation sector can be repeated for the
matter sector of the Hamiltonian constraint, and in an analogous fashion,
this suggests that inverse triad operators should be introduced
based on the form of $\rho$ alone, rather than the form of
$N \mH_{\rm matter} = N \sqrt q \, \rho$.

Further ambiguities appear in the actual definition of the inverse volume
operator corresponding to $1/V$, where $V = \sqrt{p_1 p_2 p_3}$.  This can
be seen by the fact that in the quantum theory, there exists a whole family of
commutators that can be used in order to define an inverse volume operator
following the methods of \cite{Thiemann:1996aw},
\be
\wh{\, \f{1}{V} \,} = (\wh{V})^p \bigg( f(p_i) \, e^{-i y \t\mu \ga k_i / 2} \,
\Big[ e^{i y \t\mu \ga k_i}, V^s \Big] \, e^{-i y \t\mu \ga k_i / 2}
\bigg)^{t(1+p)},
\ee
where $\t\mu(p_i)$ is in general some function of the $p_i$, and the specific
values for the exponent $t$ and the function $f(p_i)$ depend on the functional
form of $\t\mu$, as well as the numerical values of $y$ and $s$.  Note that
the factor-ordering has been chosen such that the operator is symmetric (the
$\h{V}$ and $f(p_i)$ commute with the composite operator to their right).

While this form of the inverse triad operator may seem a little opaque,
it can be rewritten in the following simpler form,
\be
\wh{\, \f{1}{V} \,} = (\wh{V})^p \bigg( \f{1}{2 s \alpha(p_i)} \Big[
|\h{V}^r + \alpha(p_i)|^s - |\h{V}^r - \alpha(p_i)|^s \Big]
\bigg)^{(1+p)/r(1-s)},
\ee
where $r$ depends on the functional dependence of $\t\mu$ on the $p_i$,
and $\alpha(p_i)$ depends on the choice of $y$ and $\t\mu$.  Note that
the four parameters are constrained by $\alpha(p_i) > 0$, $p > 0$,
$r > 0$ and $0 < s < 1$.  Despite these restrictions on the parameters,
it is clear that there is a large family of possible inverse triad operators
for $\wh{V^{-1}}$.

Let us verify that these are all reasonable inverse volume operators:
assuming that this operator acts on a state that is an eigenvector of $\h V$
with eigenvalue $V$, it is easy to check that for large $V \gg \alpha(p_i)$
the righthand side of the equation provides an excellent approximation to
$1/V$.  However, for small $V$ the eigenvalue of $\wh{V^{-1}}$ is significantly
different from $1/V$, and in particular the eigenvalue of $\wh{V^{-1}}$
vanishes when $V = 0$. Thus, the inverse volume operator annihilates zero
volume states, rather than diverging as $1/V$.

Some of the ambiguities can be resolved by the following argument:
as the holonomies appearing in the field strength operator of the
Hamiltonian constraint operator are evaluated on lengths that are
multiples of $\b\mu_i$, it seems reasonable to assume that the
holonomies appearing in the inverse triad operators should also
have a length that is a multiple of $\b\mu_i$.  In this case, as
\be
e^{-i y \b\mu_1 \ga k_1 } f(V) = f(V + 4 \pi \ga \lp^2 \sqrt \Delta y),
\ee
$r = 1$ is naturally selected, and $\alpha(p_i) = \alpha$ in this case
is independent of the $p_i$.  However, even with this simplification,
three ambiguities remain to be fixed:
\be \label{inv-v}
\wh{\, \f{1}{V} \,} = (\wh{V})^p \bigg( \f{1}{2 s \alpha} \Big[
|\h{V} + \alpha|^s - |\h{V} - \alpha|^s \Big]
\bigg)^{(1+p)/(1-s)},
\ee

One of the most common inverse triad operators in LQC, often denoted by $B(v)$,
was given earlier in the `F' loop quantization of the closed FLRW model in
\eqref{bv1}.  This particular inverse volume operator is obtained by setting
$\alpha = 2 \pi \ga \lp^2 \sqrt\Delta$, $p=1$ and $s=1/3$ in \eqref{inv-v},
giving (recall that $V = 2 \pi \ga \lp^2 \sqrt\Delta |v|$)
\be \label{bv2}
B(v) = \f{|v|}{2 \pi \ga \lp^2 \sqrt\Delta} \bigg( \f{3}{2}
\Big[ |v+1|^{1/3} - |v - 1|^{1/3} \Big] \bigg)^3.
\ee
This particular inverse triad operator is defined via the $\b\mu_i$ holonomies
that arise naturally in LQC, and therefore we have $r = 1$ and constant
$\alpha$, as expected from the arguments given above.

In addition, while this is not an ambiguity, it is important to point out that inverse
volume corrections vanish in non-compact space-times.  In a non-compact space-time,
it is necessary to introduce a fiducial cell $\mV$ as a regulator in order to
ensure that integrals do not diverge, and then $V$ corresponds to the physical
volume of $\mV$.  However, in the quantum theory, the inverse triad operators
depend on the choice of $\mV$ in a different way than $1/V$ does in the
classical theory, as can be checked explicitly in these equations.  This is
not surprising as operators often depend on regulators, and the correct
way to deal with this is to remove the regulator, namely to take the limit
as $\mV$ approaches $\mathbb{R}^3$.  In this limit, the eigenvalues $V$
become larger and larger (even at the bounce point), and so the difference
between $1/V$ and the eigenvalue of \eqref{inv-v} vanishes.  Therefore,
there are no effects from inverse triad operators in the effective theory
for non-compact spaces in LQC%
\footnote{As pointed out in \cite{Bojowald:2011iq},
the fact that inverse triad corrections
vanish for non-compact spaces may indicate that
they are not being implemented correctly in LQC.
However, as far as the authors are aware, there
there do not exist any well-defined inverse
triad operators in LQC that do not vanish
in non-compact spaces.}.
This is clearly not the case for compact spaces though, like the Bianchi
IX cosmology considered here that has the topology $\mathbb{S}^3$.
Therefore, inverse triad corrections will appear in the effective
Hamiltonian and their strength will be proportional to $\lp^3/V$.

Finally, there is also another possible definition for inverse
triad operators that is worth considering, namely
\be \label{trivial}
\wh {\, \f{1}{V} \,} \: | \Psi_V \ket = \begin{cases}
0 & \text{if  } \h{V} | \Psi_V \ket = 0,\\
\left( \h{V} \right)^{-1} \: | \Psi_V \ket & \text{otherwise.}\\
\end{cases}
\ee
where the notation $| \Psi_V \ket$ indicates that the state is an eigenstate
of the $\h{V}$ operator.  This operator was first used in the constraint
operators of lattice LQC in order to ensure that the constraints weakly
commute \cite{WilsonEwing:2012bx}.  Note that if the inverse triad operators
in lattice LQC were defined following \eqref{inv-v}, then the constraints
would no longer weakly commute.

This is an attractive choice for two reasons, namely that its action is
particularly simple, and it avoids all of the ambiguities that are
present in \eqref{inv-v}.  It is for these reasons that we choose to define
the inverse triad operators in Sec.\ \ref{ss.k-b-ix-ch} in this manner.

Furthermore, the choice \eqref{trivial} is also possible in the full theory,
where one of its advantages is that it would considerably simplify calculations.
In addition, it has been pointed out that in some relatively simple models,
implementing inverse triad operators via commutators leads to the incorrect
quantum theory \cite{Livine:2013zha}.  This last result indicates that inverse
volume operators like \eqref{inv-v} may not give the correct quantum theory,
perhaps the operator \eqref{trivial} is a better choice?

\subsection{Ambiguities in More General Inverse Triad Operators}
\label{ss.amb-triad}

In addition to the ambiguities concerning the definition of the inverse volume
operator, in anisotropic space-times one must define other more general inverse triad
operators.  Once again, there exists a choice in how to group different terms.
For example, there are several operators that could represent a term of the
form $(p_1 p_2)^{-1}$, including the three possibilities
\be \label{choices}
\wh{\f{1}{p_1 p_2}}, \quad
\wh{\, \f{1}{p_1} \,} \times \wh{\, \f{1}{p_2} \,}, \quad
\wh{p_3} \times \left( \wh{\, \f{1}{V} \,} \right)^2,
\ee
among others, which are \emph{a priori} inequivalent. (Note that all of these
operators commute so the ordering is not important.) While the form of
$\wh{V^{-1}}$ is given above (containing ambiguities), the first two
possibilities listed here remain to be defined.  One might initially choose
to define something like
\be
\wh{\, \f{1}{x} \,} = \bigg( \f{1}{2 s \alpha} \Big[
(\h{x}^r + \alpha)^s - (\h{x}^r - \alpha)^s \Big] \bigg)^{1/r(1-s)},
\ee
with $x = p_1 p_2$ or $x = p_i$, but such an operator cannot be generated by inverse
triad operators constructed from the $\b\mu_i$-type holonomies that are typically
used in LQC.  Instead, inverse triad operators defined via $\b\mu_i$ holonomies are
of the form \cite{Ashtekar:2009um}
\be \label{inv-triad}
\wh{\, \f{1}{x} \,} = \wh{ \left( \f{V^q}{x} \right) }
\times \left( \wh{\, \f{1}{V} \,} \right)^q,
\ee
where this holds for all $x$, and in particular for $x = p_1 p_2$ or $x = p_i$.
Note that $q$ must be taken to be sufficiently large so that the operator on the
left does not contain any inverse powers of a triad.  This result can be derived
simply by seeing how $\b\mu_i$ holonomies act on the $p_i$,
\be
e^{-i y \b\mu_1 \ga k_1 } \, \Big( p_1^n \Big) \, e^{i y \b\mu_1 \ga k_1 } \, | \Psi \ket
= p_1^n \cdot \left( \f{V + 4 \pi \ga \lp^2 \sqrt \Delta y}{V} \right)^{2n} \,
| \Psi \ket,
\ee
where we have assumed $n > 0$ and $V \neq 0$.  It is easy to see how this can
be generalized to products of $p_i$ raised to arbitrary positive powers.  As
this gives a shift in terms of the volume rather than the $p_i$, it turns out
that any inverse triad operator built from $\b\mu_i$ holonomies will in fact
be of the form \eqref{inv-triad}.

What is particularly nice about this result is that any inverse triad operator
is expressed in terms of the original inverse volume operator.
Furthermore, it is clear that the nature of the ambiguity $q$ is very similar to the
choice of $p$ in the inverse volume operator \eqref{inv-v}, and in fact the two combine
by addition when the full inverse triad operator is written out in full,
\be \label{inv-t}
\wh{\, \f{1}{x} \,} = \wh{ \left( \f{V^{p+q}}{x} \right) }
\bigg( \f{1}{2 s \alpha} \Big[ |\h{V} + \alpha|^s - |\h{V} - \alpha|^s \Big]
\bigg)^{(1+p+q)/(1-s)},
\ee
Thus, if one chooses to construct inverse triad operators by using the family of
$\b\mu_i$ holonomy operators, all three options in \eqref{choices} are in fact
equivalent, modulo the choice of $q$ in each operator, which may in principle be
different.  It is important to keep in mind that operators with a different choice
of $q$ vary by factors of
\be
\wh{V} \, \wh{V^{-1}} \ne \mathbb{I}.
\ee
As these two operators are not the inverse of each other, it is clear that the
quantitative predictions of LQC (at least for compact space-times) will depend
on the specific choices made here, namely the definition of the inverse volume
operator and the choice of $q$.  Hopefully, guidance from full LQG will determine
the correct way to proceed, but at this point there remain many ambiguities.

Finally, it is worth pointing out that if one chooses to define inverse triad
operators following \eqref{trivial} rather than \eqref{inv-v},
\be \label{trivial-pi}
\wh {\, \f{1}{p_i} \,} \: | p_1, p_2, p_3 \ket = \begin{cases}
0 & \text{if  } \h{p_i} | p_1, p_2, p_3 \ket = 0,\\
\left( \h{p_i} \right)^{-1} \: | p_1, p_2, p_3 \ket & \text{otherwise,}\\
\end{cases}
\ee
then all of the ambiguities in \eqref{inv-t} are resolved.  The fact that no
ambiguities appear in inverse triad operators of
the form \eqref{trivial-pi} is a very appealing
property, and provides an argument in its support as a potential inverse triad
operator.  On the other hand, if inverse triad effects are expected to be
important in the effective theory, then \eqref{trivial-pi} is not viable, as
it does not allow for any inverse triad corrections in the effective Hamiltonian,
whether the space is compact or non-compact.  So, in order to choose between the
possibilities of \eqref{inv-t} and \eqref{trivial-pi}, one must decide whether
inverse triad corrections are expected to be important in the effective
theory, or whether it is best to avoid the large number of ambiguities plaguing
the definition of the standard inverse triad operators.

In summary, in order to fully resolve the ambiguities concerning inverse triad
operators in LQC, it seems necessary to await a better understanding of inverse
triad operators in full LQG, where a number of ambiguities also remain to be
resolved.  Until then, while it will be interesting to study the various possible
effects of inverse triad operators and perhaps obtain some insights for the full
theory, it will be necessary to remain aware of the limitations of these results
due to the ambiguities in their definition.

\subsection{A Bound on the Matter Energy Density?}
\label{ss.rho}

We now discuss bounds on the matter energy density.  Recall that in anisotropic models,
the energy density $\rho$ does not capture the full space-time curvature since the
Weyl curvature is non-zero, and therefore the expansion and shear scalars are more
important.  Nonetheless, it is of interest to determine whether it is possible
to obtain a bound on the energy density in addition to the bounds on $\theta$
and $\sigma^2$ that have already been found.

In this section, we attempt to derive such a bound for the energy density on the
constraint surface.  It is important to keep in mind that this is a
strong requirement, and it is possible that the energy density may be bounded
dynamically even if there is no general bound on the constraint surface.  For
example, this is precisely what happens in the `F' loop quantization of the
closed FLRW model where $\rho$ is not bounded on the constraint surface,
but numerical simulations show that for macroscopic universes
$\rho_{\rm max} \approx 0.41 \rho_{\rm Pl}$ \cite{Ashtekar:2006es}.

As discussed in the Introduction, the issue of a bound for the energy density on
the constraint surface is subtle in spatially curved models.  In the `A' quantization
of the Bianchi II model, it is necessary to impose the weak energy condition, while
in the `A' loop quantization of the Bianchi IX model, inverse triad corrections
are needed to obtain bounds on the energy density \cite{Corichi:2009pp}.
Similarly, in the `A' loop quantization of the closed FLRW model, inverse triad
corrections are again necessary in order to bound $\rho$ \cite{Corichi:2013usa}.
These results lead us to consider the possibility that inverse triad effects may
bound the energy density in the `K' loop quantization of the Bianchi IX model
as well.

Of course, as we just saw, there are many ambiguities in the definition of inverse
triad operators in LQC, and so any discussion of their effects will necessarily be
qualitative.  We must also keep in mind that this discussion is at the effective
level, with all the usual caveats this entails.

Let us begin by recalling the fact that in spatially flat homogeneous cosmologies,
the energy density is bounded everywhere on the constraint surface in the effective
theory once holonomy effects are included.  This is not the case in the
presence of spatial curvature; a concrete example is the closed FLRW model studied
in Sec.\ \ref{sss.cl-f}, where the effective constraint \eqref{effham1} is
\be
\mC_H^{\rm eff(F)} = - \f{3 p^{3/2}}{8 \pi G \gamma^2 \Delta} \bigg[
\sin^2 \b\mu\left(c - \f{\lo}{2}\right) - \sin^2 \bar \mu
+ (1 + \gamma^2) \f{\lo^2 \Delta}{4 p} \bigg] + p^{3/2} \rho \approx 0,
\ee
where $\rho$ is the energy density of the matter field (for example, in the case of
a massless scalar field $\rho = p_\phi^2 / 2 p^3$).  On the constraint surface,
\be
\rho = \f{3}{8 \pi G \gamma^2 \Delta} \bigg[
\sin^2 \b\mu\left(c - \f{\lo}{2}\right) - \sin^2 \bar \mu
+ (1 + \gamma^2) \f{\lo^2 \Delta}{4 p} \bigg] ~,
\ee
and it is clear that the last term in this expression diverges as $p \to 0$.  However,
since this divergence is due to an inverse power of the volume, it seems likely that
including inverse triad corrections may give a bound for $\rho$, and this expectation
is shown to be correct in \cite{Corichi:2013usa}.

It is useful to briefly recall the results of \cite{Corichi:2013usa}; although we
shall define a slightly different form of inverse volume corrections, we arrive at
the same qualitative results.  Defining inverse volume corrections to have the form
\be \label{inv-v-x}
\f{1}{V^x} \to \Big[ \sqrt{V + 1} - \sqrt{|V - 1|} \Big]^{2x},
\ee
the effective Hamiltonian \eqref{effham1} becomes
\ba
\mC_H^{\rm eff(F),inv} &=& - \f{3 p^{3/2}}{8 \pi G \gamma^2 \Delta} \bigg[
\sin^2 \b\mu\left(c - \f{\lo}{2}\right) - \sin^2 \bar \mu  \nn \\ &&
+ (1 + \gamma^2) \f{\lo^2 \Delta}{4} \cdot
\Big[ \sqrt{V + 1} - \sqrt{|V - 1|} \Big]^{4/3}\bigg]
+ p^{3/2} \rho \approx 0,
\ea
where for the case of a massless scalar field,
\be \label{rho-phi}
\rho = \f{p_\phi^2}{2} \cdot \Big[ \sqrt{V + 1} - \sqrt{|V - 1|} \Big]^4,
\ee
note that this can be generalized to be any perfect fluid.
Then, on the constraint surface
\ba
\rho &=& \f{3}{8 \pi G \gamma^2 \Delta} \bigg[
\sin^2 \b\mu\left(c - \f{\lo}{2}\right) - \sin^2 \bar \mu  
+ (1 + \gamma^2) \f{\lo^2 \Delta}{4} \cdot
\Big[ \sqrt{V + 1} - \sqrt{|V - 1|} \Big]^{4/3}\bigg] \nn \\
& \le & \f{3}{8 \pi G \gamma^2 \Delta} \bigg[
2 + (1 + \gamma^2) \f{\lo^2 \Delta}{4} \cdot
\Big[ \sqrt{V + 1} - \sqrt{|V - 1|} \Big]^{4/3}\bigg],
\ea
which is bounded, as the function $\sqrt{V+1} - \sqrt{|V-1|}$ reaches its maximum
at $V=1$ with the value $\sqrt{2}$.  The particular upper bound of the energy
density is not particularly important as this bound is quite sensitive to the exact
form of the inverse triad corrections (e.g., compare the bound given here to the
one in \cite{Corichi:2013usa}), but the important result is the existence of a bound.

Now we will consider the possibility that inverse triad operators may bound
the energy density in the Bianchi IX space-time as well.
In order to make the results of this investigation as clear as possible, it is
helpful to define a specific inverse triad correction in the effective Hamiltonian
constraint.  This specific choice should not be viewed as the ``correct'' form
of inverse triad corrections, but rather as one of the simplest implementations
possible that can be used in order to perform explicit calculations.  While
qualitative results should not be affected by the specific form of the inverse
triad corrections used here, any quantitative results cannot be trusted until
the ambiguities of inverse triad operators in LQC are resolved.

With these caveats in mind, the basic inverse volume correction that we shall
use in order to study their qualitative effects in the semi-classical limit is
obtained by taking $x=6$ in \eqref{inv-v-x},
\be \label{eff-inv-v}
\f{1}{V^6} \to  \Big[ \sqrt{V + 1} - \sqrt{|V - 1|} \Big]^{12}.
\ee
For the sake of simplicity, we will only work with this one particular inverse
volume correction, and therefore terms of the form $1/p_1 p_2$ for example will
be treated as
\be
\f{1}{p_1 p_2} = \f{p_3 V^4}{V^6} \to p_3 V^4 \times
\Big[ \sqrt{V + 1} - \sqrt{|V - 1|} \Big]^{12}.
\ee
This choice is of course arbitrary, and the sole reason for this particular
choice is that it will considerably simplify the form of the effective
Hamiltonian constraint.  As stated above, while other inverse triad corrections
will not give the same quantitative physics, the qualitative behaviour of
inverse triad effects should not be affected by these choices.  So long
as we are only interested in qualitative predictions, it is safe to only
consider this rather simple implementation of inverse triad corrections.

Following these choices and taking $N=1$, the effective LQC Hamiltonian for
the Bianchi IX space-time including holonomy and inverse triad corrections
is
\begin{align}
\mC_H^{\rm eff(F),inv} = & -\f{V}{8 \pi G \ga^2 \Delta} \Big[ \sin \b\mu_1 \ga k_1
\sin \b\mu_2 \ga k_2 + \sin \b\mu_2 \ga k_2 \sin \b\mu_3 \ga k_3
+ \sin \b\mu_3 \ga k_3 \sin \b\mu_1 \ga k_1 \Big] \nn \\ &
- \f{\lo^2 V}{32 \pi G} \Big[ 2 V^4 (p_1^2 + p_2^2 + p_3^2)
- p_1^4 p_2^4 - p_2^4 p_3^4 - p_3^4 p_1^4 \Big]
\cdot \Big[ \sqrt{V + 1} - \sqrt{|V - 1|} \Big]^{12} \nn \\ &
+ V \rho \approx 0,
\end{align}
where $\rho$ is the matter energy density, which when expressed in
terms of phase space variables may include inverse triad corrections
of its own as in \eqref{rho-phi}.

Clearly, the specific form of the inverse triad corrections comes from the
simple definition given in this section, as well as the choice to implement
inverse triad corrections on ${}^{(3)}\!R$ rather than $\sqrt q \: {}^{(3)}\!R$,
as explained at the beginning of Sec.\ \ref{ss.amb-inv}.  A nice property of
the ensuing effective Hamiltonian is that each of its terms includes the same
prefactor $V$, coming from $\sqrt q$.

On the constraint surface of this effective Hamiltonian,
\begin{align}
\rho = & \f{1}{8 \pi G \ga^2 \Delta} \Big[ \sin \b\mu_1 \ga k_1
\sin \b\mu_2 \ga k_2 + \sin \b\mu_2 \ga k_2 \sin \b\mu_3 \ga k_3
+ \sin \b\mu_3 \ga k_3 \sin \b\mu_1 \ga k_1 \Big] \nn \\ &
+ \f{\lo^2}{32 \pi G} \Big[ 2 V^4 (p_1^2 + p_2^2 + p_3^2)
- p_1^4 p_2^4 - p_2^4 p_3^4 - p_3^4 p_1^4 \Big]
\cdot \Big[ \sqrt{V + 1} - \sqrt{|V - 1|} \Big]^{12}.
\end{align}
The terms on the first line are clearly bounded, so if the terms on the second
line can be bounded as well, then it follows that $\rho$ is also bounded.

In the classical theory, one of the sources of divergences in the terms on
the second line are the denominators blowing up as $p_i \to 0$.  Now that
inverse triad corrections have been included, this is no longer possible.
Therefore, if all three $p_i$ are approaching zero, then the matter energy
density is automatically bounded by the constraint $\mC_H = 0$.  (Of course, the
actual numerical bound will depend on the specific definition of the inverse
triad operator.)

However, there is a second possible type of divergence that occurs in
anisotropic space-times in general relativity that are called cigar-like
singularities.  In a cigar-like singularity, one of the scale factors
diverges while the other two go to zero.  In this case, some of the
terms on the second line can diverge due to large numerators
rather than small denominators.  For example, if we
take $p_1 \to \infty, p_2 \to 0$ and $p_3 \to 0$ such that $V \sim 1$,
the first term on the second line diverges.  While inverse triad corrections
prevent any divergences occurring due to denominators vanishing, they cannot
provide any protection against numerators that become large.

Therefore, it is clear that the constraint $\mC_H = 0$ alone is not sufficient
to derive a bound for $\rho$.  Instead, in order to see whether it is possible
to bound $\rho$, it will be necessary to study the dynamics in order to
determine exactly how fast the growing scale factor diverges and whether
it is kept in check by the other scale factors going to zero.

Despite the fact that $\mC_H = 0$ by itself is not sufficient to bound $\rho$,
it seems that the matter energy density will in fact be dynamically bounded.
This can be seen as follows: for one of the scale factors to diverge while
ensuring that $V$ remains small, at least one of the other two scale factors
must vanish.  However, LQC effects are expected to become important when the
space-time curvature becomes large, and this is precisely what will happen
in this setting where the shear is necessarily very large.  Once the space-time
curvature becomes large, LQC effects are expected to become important and cause
a bounce, at which point the two decreasing scale factors start to grow and
the originally growing scale factor will start to shrink.  Then, as any scale
factor will necessarily remain bounded above (and below), it follows that
$\rho$ must remain bounded at all times.

Therefore, while the relation $\mC_H = 0$ alone cannot bound $\rho$ in the
Bianchi IX space-time ---even when inverse triad effects are included---
the presence of a bounce is expected to dynamically bound the matter
energy density.

\section{Conclusions}
\label{s.disc}

The quantum geometry effects that arise in loop quantum cosmology come from
field strength and inverse triad operators.  In this paper, we have studied
a class of ambiguities that arise in the definition of each of these types of
operators.

In the simplest models, the field strength operator is determined by expressing
$F_{ab}{}^k$ in terms of holonomies of the Ashtekar-Barbero connection around
a closed loop, and quantizing the resulting expression.  This procedure is
expected to be valid in the settings where it can be implemented, but in the
open FLRW and spatially curved Bianchi models the resulting expression for the
field strength is not an almost periodic function of the connection, and so it
is not known how to quantize it.  Therefore, in these models an ambiguity
arises as it is necessary to consider alternative prescriptions.  There are two
main proposals: the `K' loop quantization where one constructs operators
corresponding to the parallel transport of the extrinsic curvature along open
edges, and the `A' loop quantization where one works with holonomies of the
Ashtekar-Barbero connection evaluated along open edges.

At first, it might seem that the `A' loop quantization is a better choice
as it is constructed from holonomies of $A_a^i$, one of the elementary variables
of LQG, but a closer examination shows that the `K' loop quantization does
a significantly better job of capturing the physics of the LQC of the closed FLRW
model.  Motivated by this observation, we construct the `K' loop quantization
of the Bianchi IX model and study its effective theory.  We find that in the
`K' effective theory, the expansion and shear scalars are bounded above due
to quantum geometry effects.  Since the boundedness of the expansion and
shear scalars plays an important role in the avoidance of singularities, this
result is a strong indication that singularities are resolved in the effective
theory.  Analogous results are obtained for the Bianchi II model in Appendix
\ref{s.b-ii}.  These results are nicely concordant with those of the LQC
of Bianchi I, where the effective theory also bounds the expansion and shear
scalars; what is more, the numerical values of these bounds for all of the Bianchi
models are in exact agreement.  Thus, we find that the `K' loop quantization
is closer to the standard loop quantization than the `A' loop quantization,
for both FLRW and Bianchi models.

There are also ambiguities in the definition of inverse triad operators.
One of the first problems is to determine how the terms in the classical
constraint should be grouped together, namely whether we should allow
factors in the numerator to cancel terms in the denominator before
quantization or not.  Further ambiguities arise as inverse triad operators
are typically defined via commutators, and there exists a large family of
commutator operators that give the same large volume limit, however these
operators can be considerably different in the small volume regime.
Therefore, it is important to determine which particular inverse triad
operator is the correct one.

We show that in LQC, if we define the inverse triad operators via
$\b\mu_i$ holonomies, there are three ambiguities that need to be fixed.
Since there is no guarantee that all types of inverse triad operators
are constructed in the same manner, we cannot assume that the three
ambiguities will be resolved in the same manner in all inverse triad
operators.  These problems can be avoided by defining an alternative
trivial inverse triad operator, given in Eq.\ \eqref{trivial}, which
can also be used in the full theory.  The advantages of this particular
inverse triad operator is that it has a particularly simple action and
no ambiguities arise in its definition.

We close with some words regarding the dynamics of the effective theory.
The existence of upper bounds for the expansion and shear scalars, together
with the presence of a bounce in the LQC of the other homogeneous space-times
that have been studied so far, strongly suggests that a bounce will occur
in the LQC of the Bianchi IX model.  It would be interesting to study the
effective dynamics in order to determine whether this is indeed the case,
and also address another important open question regarding the approach
to the high curvature regime: classically, it is known that the approach
to the Bianchi IX singularity is chaotic \cite{Barrow:1981sx}, but it
has been suggested that the chaotic behaviour may disappear in LQC
\cite{Bojowald:2003xe,Bojowald:2003xf,Bojowald:2004ra}.  A detailed
study of the effective dynamics obtained here for the `K' loop quantization
could provide an answer to this issue; we leave this question
for future work.

\acknowledgments

The authors would like to thank
Brajesh Gupt
and
Madhavan Varadarajan
for helpful discussions.
This work is supported by  NSF grant
PHYS106874 and a grant from the John Templeton Foundation.
The opinions expressed in this publication are those of the
authors and do not necessarily reflect the views of the John
Templeton Foundation.

\begin{appendix}

\section{The Bianchi II Space-time}
\label{s.b-ii}

In this appendix, we describe the `K' loop quantization of the Bianchi II
space-time and show how the resulting effective theory has the same bounds
on the expansion and shear scalars that arise for the LQC of Bianchi I and
for the `K' loop quantization of Bianchi IX.

\subsection{The Classical Theory}
\label{ss.b-ii-c}

We start by briefly recalling the classical structure of the Bianchi II
space-time, for further details see \cite{Ashtekar:2009um}.

While the Bianchi II space-time is homogeneous, the spatial curvature is non-zero
and this is encoded in the fact that the right-invariant vector fields $\oe^a_i$
(which we shall use as fiducial triads) have a non-trivial commutator,
\be
[ \oe_i, \oe_j ]^a = -C^k_{ij} \, \oe^a_k,
\ee
where $C^1_{23} = -C^1_{32} = \t\alpha$ are the only non-zero structure constants.
Using $x, y, z$ as coordinates, the fiducial triads are given by
\be \label{bii-oe}
\oe^a_1 = \left( \f{\partial}{\partial x} \right)^a, \qquad
\oe^a_2 = \t\alpha z \left( \f{\partial}{\partial x} \right)^a
+ \left( \f{\partial}{\partial y} \right)^a, \qquad
\oe^a_3 = \left( \f{\partial}{\partial z} \right)^a.
\ee

As the space is homogeneous and non-compact, it is necessary to restrict integrals
to a finite region determined by a fiducial cell.  We take the fiducial cell to
be rectangular, whose edges (parallel to the coordinate axes) are of lengths
$L_1, L_2, L_3$ with respect to the fiducial metric $\oq_{ab} = \ow_a^i \ow_{bi}$,
where $\ow_a^i$ are the fiducial co-triads dual to $\oe^a_i$.

Then, it is possible to parametrize the Ashtekar-Barbero connection and the
densitized triads by
\be \label{bii-E}
A_a^i = \f{c_i}{L_i} \, \ow_a^i, \qquad
E^a_i = \f{p_i L_i}{L_1 L_2 L_3} \sqrt{\oq} \, \oe^a_i,
\ee
where it is understood that there is no sum over $i$.  This choice ensures
that the Gauss and diffeomorphism constraints are automatically satisfied,
and so only the Hamiltonian constraint is left, which for $N=1$ is given by
\cite{Ashtekar:2009um}
\begin{align}
\mC_H = & \, -\f{1}{8 \pi G \ga^2 \sqrt{p_1 p_2 p_3}}
\Bigg[ p_1 p_2 c_1 c_2 + p_2 p_3 c_2 c_3 + p_3 p_1 c_3 c_1 \nn \\ & \qquad
+ \alpha p_2 p_3 c_1 - (1 + \ga^2) \left( \f{\alpha p_2 p_3}{2 p_1} \right)^2 \Bigg]
+ \f{p_\phi^2}{2 \sqrt{p_1 p_2 p_3}} \approx 0,
\end{align}
where $\alpha = L_2 L_3 \t\alpha / L_1$.  The non-zero Poisson brackets for
the gravitational sector are
\be
\{ c_i, p_j \} = 8 \pi G \ga \de_{ij},
\ee
just as for the other Bianchi cosmlogies.

In order to determine the `K' loop quantization, it is necessary to rewrite these
results in terms of the extrinsic curvature
\be
K_a^i = \f{1}{\ga} \Big( A_a^i - \Gamma_a^i \Big) = k^i \ow_a^i,
\ee
where there is no sum over $i$ in the last term, and the spin-connection
$\Gamma_a^i$ can be calculated from the densitized triads, (see
\cite{Ashtekar:2009um} for a derivation of this result)
\be
\Gamma_a^1 = \f{\alpha p_2 p_3}{2 p_1^2} \f{\ow_a^1}{L_1}, \qquad
\Gamma_a^2 = -\f{\alpha p_3}{2 p_1} \f{\ow_a^2}{L_2}, \qquad
\Gamma_a^3 = -\f{\alpha p_2}{2 p_1} \f{\ow_a^3}{L_3}.
\ee

It is easy to check that the Hamiltonian constraint, again for $N=1$, becomes
\be
\mC_H = -\f{1}{8 \pi G \sqrt{p_1 p_2 p_3}}
\Bigg[ p_1 p_2 k_1 k_2 + p_2 p_3 k_2 k_3 + p_3 p_1 k_3 k_1
- \left( \f{\alpha p_2 p_3}{2 p_1} \right)^2 \Bigg]
+ \f{p_\phi^2}{2 \sqrt{p_1 p_2 p_3}} \approx 0,
\ee
while the non-zero Poisson brackets are
\be
\{ k_i, p_j \} = 8 \pi G \de_{ij}.
\ee

\subsection{The `K' Loop Quantum Cosmology of the Bianchi II Model}
\label{ss.b-ii-q}

The `K' loop quantization of the Bianchi II model follows essentially the same steps
as in Secs.\ \ref{ss.k-b-ix-kin} and \ref{ss.k-b-ix-ch}, with only one difference:
the multiplicative term in the Hamiltonian constraint operator changes as a result
of the Hamiltonian constraint being different.  Indeed, the kinematical Hilbert
space is exactly the same as the one defined in Sec.\ \ref{ss.k-b-ix-kin}, and the
extrinsic curvature is once again expressed in terms of complex exponentials following
\eqref{k-hat}.

Therefore, the only difference appears in the Hamiltonian constraint operator,
\be
\hbar^2 \partial_\phi^2 \Psi =
\f{\hbar V}{16 \ga \sqrt\Delta} \Big[
\sqrt{v(v+4)} \Psi^+_4 - v \Psi^+_0
- v \Psi^-_0 + \sqrt{v(v-4)} \Psi^-_4 \Big]
+ \f{1}{4 \pi G} \left( \f{\alpha p_2 p_3}{2 p_1} \right)^2 \Psi,
\ee
where the last term in the operator is different from the Bianchi IX case. As before, the
$\Psi^\pm_{0,4}$ are given in \eqref{psi-pm}, and the inverse triad operators are defined
to be \eqref{inv-v-app}, just as for the Bianchi IX model.  Once again, other choices for
the definition of inverse triad operators are available; here we went with the simplest
possibility.

\subsection{The Effective Theory of the `K' Loop Quantization}
\label{ss.b-ii-e}

The effective theory for the `K' loop quantization of the Bianchi II space-time
obtained from the effective Hamiltonian constraint is, for $N=1$,
\begin{align}
\mC_H = & \, -\f{\sqrt{p_1 p_2 p_3}}{8 \pi G \ga^2 \Delta}
\Big[ \sin \b\mu_1 \ga k_1 \sin \b\mu_2 \ga k_2
+ \sin \b\mu_2 \ga k_2 \sin \b\mu_3 \ga k_3
+ \sin \b\mu_3 \ga k_3 \sin \b\mu_1 \ga k_1 \Big] \nn \\ & \quad
+ \f{1}{8 \pi G \sqrt{p_1 p_2 p_3}} \left( \f{\alpha p_2 p_3}{2 p_1} \right)^2
+ \f{p_\phi^2}{2 \sqrt{p_1 p_2 p_3}} \approx 0.
\end{align}
Here we are including holonomy corrections, but not inverse triad effects.
It is a straightforward calculation to show that
\be
\d p_1 = \f{p_1}{\ga \sqrt\Delta} \Big( \sin \b\mu_2 \ga k_2
+ \sin \b\mu_3 \ga k_3 \Big) \cos \b\mu_1 \ga k_1,
\ee
which is exactly the same result as the one obtained in the effective theory
for the `K' loop quantization of the Bianchi IX space-time, given in \eqref{p1-dot}.

As the directional Hubble rates in the Bianchi II space-time are given by, for
example,
\be
H_1 = \f{\d p_2}{2 p_2} + \f{\d p_3}{2 p_3}
- \f{\d p_1}{2 p_1},
\ee
just as in the Bianchi I and Bianchi IX space-times, it follows that the
expansion
\be
\theta = \f{1}{3} (H_1 + H_2 + H_3)
\le \f{3}{2 \ga \sqrt\De},
\ee
is always bounded, as is the shear
\be
\sigma^2 = \f{1}{3} \Big[ (H_1 - H_2)^2 + (H_2 - H_3)^2 + (H_3 - H_1)^2 \Big]
\le \f{10.125}{3 \ga^2 \De}.
\ee
The existence of an upper bound on the expansion and the shear scalars is a
strong indication that geodesic singularities may be resolved in the effective
theory for the `K' loop quantization of the Bianchi II cosmology.

In addition, a particularly nice feature of the effective theories for the `K'
loop quantization of the Bianchi type II and type IX space-times is that one obtains
identical upper bounds for the expansion and the shear that are in exact agreement
with the bounds obtained in the standard `F' loop quantization of the Bianchi I model.

\end{appendix}



\end{document}